\documentclass[PAPER, atlasdraft=false, cernpreprint, UKenglish, texmf, orcidlogo]{atlasdoc}
\usepackage{atlaspackage}
\usepackage{atlasbiblatex}

\usepackage{atlasphysics}[hepprocess=true,snippets=true] %
\graphicspath{{logos/}{figures/}}

\usepackage{ANA-TAUP-2023-01-PAPER-defs}
\usepackage{multirow}
\usepackage{subfig}

\AtlasTitle{Estimation of backgrounds from jets misidentified as $\tau$-leptons using the Universal Fake Factor method with the ATLAS detector}

\AtlasAbstract{%
Processes with $\tau$-leptons in the final state are important for Standard Model
measurements and searches for physics beyond the Standard Model.
The ATLAS experiment at the Large Hadron Collider observes $\tau$-leptons produced in proton--proton collisions only
through their decay products.
Data analyses involving hadronically decaying $\tau$-leptons face challenges due to backgrounds
from jets misidentified as $\tau$-leptons that are not modelled reliably by Monte Carlo simulations.
Data-driven methods such as the fake-factor method allow such misidentified backgrounds to be predicted by measuring
transfer factors, known as fake factors, in data from dedicated regions.
This paper describes a refined technique for determining the fake factors, the Universal Fake Factor method.
It evaluates the fake factors for a signal region by using
fake factors from samples enriched in different sources of jets misidentified as $\tau$-leptons (light-quark, gluon, $b$-quark, and pile-up jets).
Each fake factor is calculated as a linear combination of fake factors measured in these different enriched samples.
For the full Run~2 data set, the systematic uncertainty of the calculated fake factors,
evaluated using $W(\mu\nu)$ enriched event sample,
ranges from 15\% to 35\% depending
on the $\tau$-lepton's transverse momentum and charged-particle decay multiplicity.
}

\AtlasRefCode{TAUP-2023-01}

\PreprintIdNumber{CERN-EP-2025-003}

\AtlasJournal{EPJC}
\AtlasJournalRef{Eur. Phys. J. C 85 (2025) 1441}
\AtlasDOI{10.1140/epjc/s10052-025-14916-1}

\AtlasCoverEgroupAnalysisTeam{atlas-taup-2023-01-analysis-team@cern.ch}

\hypersetup{pdftitle={ATLAS document},pdfauthor={The ATLAS Collaboration}}

\begin{document}

\maketitle

\tableofcontents

%
%

%

\section{Introduction}
\label{sec:intro}

Due to their short lifetime, \tl-leptons produced in proton--proton (\pp)
collisions at the Large Hadron Collider (LHC) can be observed in the ATLAS detector only in the form of their decay products.
The \tl-lepton decays either leptonically (\taulep) or hadronically (\tauhad), with corresponding branching fractions of 35\% and 65\%.
Because neutrinos are not detected, only hadrons can be reconstructed as visible \tauhad decay products (\tauhadvis).
In analyses with \tauhadvis in the final state, jets misidentified as \tauhadvis (fakes) constitute a sizeable source of background.
Monte Carlo (MC) generators cannot properly model the \enquote{fake} background
because such modelling requires detailed simulation of soft jet properties and the detector's response to hadrons.
It is also challenging to generate enough events with misreconstructed \tauhadvis from various sources to match the integrated luminosity of the data.
Data-driven methods are therefore preferred.
This paper describes and generalizes the fake-factor (FF) method~\cite{LEHMANN2023168376} that has been applied successfully in many ATLAS and CMS
analyses~\cite{HIGG-2022-07,HIGG-2019-09,HDBS-2019-27,HIGG-2018-20,HIGG-2019-10,HIGG-2016-12,HIGG-2017-08,EXOT-2020-18,EXOT-2022-39,EXOT-2021-15,EXOT-2020-07,CMS-HIG-21-017,CMS-EXO-22-007,CMS-EXO-21-009}.
The FF method has also been used to estimate the background
due to jets misidentified as electrons or muons~\cite{EGAM-2019-01}.

The FF method estimates the fake background by measuring the transfer factors
that allow the background in a signal region (SR) to be predicted by using data from a dedicated control region (CR).
In this paper, the only difference between the SR and the CR is a single selection criterion: the \tauhadvis identification (ID), as described in Sec.~\ref{sec:eventreco}, which is inverted in the CR definition.
Due to this similarity, the CR is treated as part of the SR and referred to as the antiID subregion of the SR.
Its complement, where the \tauhadvis ID requirement is satisfied, is called the ID subregion of the SR.
The transfer factors used to estimate the fake background are referred to as fake factors and are determined in dedicated regions, referred to as Determination Regions (DRs) in this paper.

In general, several types of jets are sources of fake \tauhadvis.
Four main sources of fake \tauhadvis are considered: jets originating from
a $b$-quark (\bfake), any lighter quark flavour (\qfake), a gluon (\gfake),
or \pileup (\pfake fakes), i.e.\ multiple \pp interactions
in the same and neighbouring bunch crossings.
Jets originating from a $c$-quark are not treated as a separate category,
as they are expected to form a small fraction of the total fakes in typical physics analyses.
For example, the fraction of $c$-quark originated fake \tauhadvis is below 15\% in all regions presented in this paper.
The selected types of jets have different properties, and therefore they have different probabilities
to be misidentified as a \tauhadvis.
The \ff is different for each fake type, but generally it is not possible to measure the \ff
for each fake type separately because every region of phase space contains a mixture of different types.
An ideal DR for the \ff measurement would contain a sample of fakes with the same \qgbp composition
as in the SR, which is difficult to achieve.
A solution to this problem is the Universal Fake Factor (UFF) method described in this paper.
This method proposes that the FF for a particular SR can be obtained by knowing the
fractions of \qgbp fakes in the antiID subregion of the SR and using the FFs measured in samples enriched
in each type of fake (\qgbp).
The combined FF suitable for the SR is estimated as a linear combination of four FFs measured in the four
fake-enriched regions.
The combination coefficients are determined using a template fit.
The dominant systematic uncertainty comes from differences in the \ffs for different fake types,
and it is estimated using MC simulation.
The UFF method has one more advantage over more common ATLAS implementations of the FF method.
The latter typically require many CRs or DRs,
and an exhaustive validation of the predictions is often not possible.
The UFF method does not require the analysis to define dedicated DRs to measure the FFs,
as it defines its own fake-enriched regions for the FF measurement.
Instead, regions that the analysis would define to measure the FFs
can be used to validate the UFF-based fake-background prediction.

Preliminary versions of the UFF method were successfully used for
fake-background estimation in three published ATLAS analyses.
One searches for charged Higgs bosons in the \tauhad{}\,+\,jets and \tauhad{}\,+\,lepton final states~\cite{HIGG-2016-11},
and the second for dark matter produced in association with a Higgs boson decaying into \tl-leptons~\cite{HDBS-2018-50}.
The third result measures the high-mass $\tau^+ \tau^-$ production cross-section~\cite{EXOT-2022-42}.
The CMS Collaboration has also developed and successfully applied a similar method~\cite{CMS-HIG-15-007,CMS-HIG-23-003,CMS-HIG-21-001,CMS-HIG-17-020,CMS-EXO-19-016}.

The present version of the UFF method does not consider electrons, muons or photons misidentified as \tauhadvis.
These objects typically generate less background compared to the misidentified jets.

This paper is organized as follows.
The ATLAS detector is described in Section~\ref{sec:detector}.
Section~\ref{sec:datamc} introduces the data sets used in the present publication.
Details of the reconstruction and identification of  physical objects are provided in Section~\ref{sec:eventreco}.
After a summary of the FF method in Section~\ref{sec:ff}, the UFF method is described in Section~\ref{sec:uff}.
The UFF method is tested in Section~\ref{sec:validation},
its applicability and limitations are discussed in Section~\ref{sec:proscons},
and the conclusions are provided in Section~\ref{sec:conclusion}.


%

%

%
\newcommand{\AtlasCoordFootnote}{%
ATLAS uses a right-handed coordinate system with its origin at the nominal interaction point (IP)
in the centre of the detector and the \(z\)-axis along the beam pipe.
The \(x\)-axis points from the IP to the centre of the LHC ring,
and the \(y\)-axis points upwards.
Polar coordinates \((r,\phi)\) are used in the transverse plane,
\(\phi\) being the azimuthal angle around the \(z\)-axis.
The pseudorapidity is defined in terms of the polar angle \(\theta\) as \(\eta = -\ln \tan(\theta/2)\) and is equal to the rapidity
\(y = (1/2) \ln[(E + p_z)/(E - p_z)]\) in the relativistic limit.
Angular distance is measured in units of \(\Delta R \equiv \sqrt{(\Delta y)^{2} + (\Delta\phi)^{2}}\).
Transverse momentum, \pt, is the magnitude of the momentum component in the transverse plane.
}

\section{ATLAS detector}
\label{sec:detector}

The ATLAS detector~\cite{PERF-2007-01} at the LHC covers nearly the entire solid angle around the collision point.\footnote{\AtlasCoordFootnote}
It consists of an inner tracking detector surrounded by a thin superconducting solenoid, electromagnetic and hadronic calorimeters,
and a muon spectrometer incorporating three large superconducting air-core toroidal magnets.

The inner-detector system is immersed in a \qty{2}{\tesla} axial magnetic field
and provides charged-particle tracking in the range \(|\eta| < 2.5\).
The high-granularity silicon pixel detector covers the vertex region and typically provides four measurements per track,
the first hit generally being in the insertable B-layer installed before Run~2~\cite{ATLAS-TDR-19,PIX-2018-001}.
It is followed by the silicon microstrip detector, which usually provides eight measurements per track.
These silicon detectors are complemented by the transition radiation tracker (TRT),
which enables radially extended track reconstruction up to \(|\eta| = 2.0\).
The TRT also provides electron identification information
based on the fraction of hits (typically 30 in total) above a higher energy-deposit threshold corresponding to transition radiation.

The calorimeter system covers the pseudorapidity range \(|\eta| < 4.9\).
Within the region \(|\eta|< 3.2\), electromagnetic calorimetry is provided by barrel and
endcap high-granularity lead/liquid-argon (LAr) calorimeters,
with an additional thin LAr presampler covering \(|\eta| < 1.8\)
to correct for energy loss in material upstream of the calorimeters.
Hadronic calorimetry is provided by the steel/scintillator-tile calorimeter,
segmented into three barrel structures within \(|\eta| < 1.7\), and two copper/LAr hadronic endcap calorimeters.
The solid angle coverage is completed with forward copper/LAr and tungsten/LAr calorimeter modules
optimized for electromagnetic and hadronic energy measurements respectively.

The muon spectrometer (MS) comprises separate trigger and
high-precision tracking chambers measuring the deflection of muons in a magnetic field generated by the superconducting air-core toroidal magnets.
The field integral of the toroids ranges between \num{2.0} and \qty{6.0}{\tesla\metre}
across most of the detector.
Three layers of precision chambers, each consisting of layers of monitored drift tubes, cover the region \(|\eta| < 2.7\),
complemented by cathode-strip chambers in the forward region, where the background is highest.
The muon trigger system covers the range \(|\eta| < 2.4\) with resistive-plate chambers in the barrel, and thin-gap chambers in the endcap regions.

The luminosity is measured mainly by the LUCID-2~\cite{LUCID2} detector, which records Cherenkov light produced in the quartz windows of photomultipliers located close to the beampipe, complemented by measurements using the inner detector and calorimeters.
Events are selected by the first-level (L1) trigger system implemented in custom hardware,
followed by selections made by algorithms implemented in software in the high-level trigger~\cite{TRIG-2016-01}.
The L1 trigger accepts events from the \qty{40}{\MHz} bunch crossings at a rate close to \qty{100}{\kHz}, which the high-level trigger further reduces in order to record complete events to disk at about \qty{1.25}{\kHz}.
A software suite~\cite{SOFT-2022-02} is used in data simulation, in the reconstruction
and analysis of data and simulated events, in detector operations, and in the trigger and data acquisition
systems of the experiment.


%

%

\section{Data and Monte Carlo samples}
\label{sec:datamc}

The present analysis uses data that the ATLAS experiment collected under stable-beam conditions in 2015--2018 at a \textit{pp} centre-of-mass collision energy of $\sqrt{s} = 13~\tev$.
After imposing data quality requirements to ensure that the ATLAS detector was operating normally~\cite{DAPR-2018-01},
the integrated luminosity of the data sample is $140.1 \pm 1.2~\ifb$~\cite{DAPR-2021-01}.

The analysis uses two sets of MC samples called \enquote{Default} and \enquote{Powheg} in the following.
Both of them include simulated events from \ttbar, single top, and \ttV (where $V = W, Z$) production (collectively called \enquote{Top}),
quantum chromodynamics (QCD) and electroweak (EW) vector-boson production in association with jets (\vjets),
pair production of vector bosons (\VV), and dijet production.
The Default set consists of the following MC samples.
The \vjets events were generated with \SHERPA[2.2.1]~\cite{Bothmann:2019yzt}.
The \ttbar and single-top events were generated with \POWHEGBOX[v2]~\cite{Frixione:2007nw,Nason:2004rx,Frixione:2007vw,Alioli:2010xd} interfaced to \PYTHIA[8.230]~\cite{Sjostrand:2014zea}, while the \ttV events were modelled with
\MGNLO[2.3.3]~\cite{Alwall:2014hca} interfaced to \PYTHIA[8.210].
\SHERPA[2.2.1] and \SHERPA[2.2.2] were used to generate the \VV events.
The dijet events were generated using \PYTHIA[8.230].
The Powheg set differs from the Default set in two respects:
QCD \vjets events in the Powheg set were generated with \POWHEGBOX[v1]~\cite{Nason:2004rx,Frixione:2007vw,Alioli:2010xd,Alioli:2008gx}
interfaced to \PYTHIA[8.186], and dijet events were generated with \POWHEGBOX[v2] interfaced to \PYTHIA[8.245].
While the Default set provides more accurate modelling of final states with multiple jets,
the Powheg set is used to estimate systematic uncertainties.
Table~\ref{tab:mc} lists all the MC generators, the versions of parton distribution function (PDF) sets used in matrix element (ME) and
parton shower (PS) computations, the set of tuned parameters (\enquote{tune}) used for the underlying event,
and the order in the strong coupling constant \alphas to which the MEs and cross-sections are calculated.

\begin{table}
\captionsetup{aboveskip=10pt, belowskip=10pt}
\centering
\hskip-0.2cm
\caption{
List of MC generators, their settings, and the \alphas-order of the cross-section (XS) calculation.
MG5~v2.3.3 stands for \MGNLO[2.3.3].
The Sherpa, A14 and AZNLO underlying event (UE) tunes are described in Refs.~\cite{Bothmann:2019yzt}, \cite{ATL-PHYS-PUB-2014-021}, and~\cite{STDM-2012-23},
respectively.
}
\scalebox{0.8}{
\begin{tabular}{lccccccccc}
\toprule
\multirow{2}{*}{Process} & \multicolumn{2}{c}{\textbf{Generator}} && \multicolumn{2}{c}{\textbf{PDF set}} & \multirow{2}{*}{UE tune} & \multirow{2}{*}{ME order} & \multirow{2}{*}{XS order} \\
\cline{2-3}\cline{5-6}
& ME & PS && ME & PS & & & \\\hline
\vjets QCD & \multicolumn{2}{c}{\SHERPA[2.2.1]} && \multicolumn{2}{c}{\NNPDF[3.0nnlo]} & Sherpa & NLO/LO & NNLO \\\hline
\vjets EW & \multicolumn{2}{c}{\SHERPA[2.2.1]} && \multicolumn{2}{c}{\NNPDF[3.0nnlo]} & Sherpa & LO & LO \\\hline
$t\bar{t}$ & \POWHEGBOX[v2] & \PYTHIA[8.230] && \NNPDF[3.0nlo] & \NNPDF[2.3lo] & A14 & NLO & NNLO \\
Single top & \POWHEGBOX[v2] & \PYTHIA[8.230] && \NNPDF[3.0nlo] & \NNPDF[2.3lo] & A14 & NLO & NLO \\\hline
\ttV & MG5~v2.3.3 & \PYTHIA[8.210] && \NNPDF[3.0nlo] & \NNPDF[2.3lo] & A14 & NLO & NLO \\\hline
\VV & \multicolumn{2}{c}{\SHERPA[2.2.1,2.2.2]} && \multicolumn{2}{c}{\NNPDF[3.0nnlo]} & Sherpa & NLO/LO & NNLO \\\hline
Dijet & \multicolumn{2}{c}{\PYTHIA[8.230]} && \NNPDF[2.3lo] & \NNPDF[2.3lo] & A14 & LO & LO \\\hline
\vjets QCD & \POWHEGBOX[v1] & \PYTHIA[8.186] && \CT[10nlo] & CTEQ6L1 & AZNLO & NLO & NLO \\\hline
Dijet & \POWHEGBOX[v2] & \PYTHIA[8.245] && \CT[18nnlo] & \NNPDF[2.3lo] & A14 & NLO & NLO \\\bottomrule %
\end{tabular}}
\label{tab:mc}
\end{table}

MC simulation is used to model both \tl-lepton production and the production of fake \tauhadvis.
Each generated event is passed through a detailed simulation of the ATLAS detector's response to particles~\cite{SOFT-2010-01}, based on \GEANT~\cite{Agostinelli:2002hh}.
The effect of \pileup was modelled by overlaying the simulated hard-scattering event with
inelastic \(pp\) scattering events generated using \PYTHIA[8.186]~\cite{Sjostrand:2007gs}
with the \NNPDF[2.3lo] PDF set~\cite{Ball:2012cx} and the
A3 tune~\cite{ATL-PHYS-PUB-2016-017}.
Simulated events are reweighted to match the \pileup conditions observed in the data.


%

%

%
%
%
%
%
%
%
%
%
%
%
%
%
%
%
%
%
%
%
%

\section{Event reconstruction}
\label{sec:eventreco}

In the data sample used for the present analysis, the number of \textit{pp}~collisions occurring simultaneously in a bunch crossing varies from about 20 to 70.
The locations of individual \textit{pp}~collisions are called vertices, and each of them is reconstructed from the intersection of at least two tracks with $\pt > 500~\mev$.
Additional requirements on the tracks guarantee that they originate from a region where the beams overlap in the transverse plane.
The primary vertex is defined as the vertex with the largest sum of squared \pt of its matched tracks~\cite{ATL-PHYS-PUB-2015-026}.
The present analysis considers events with electrons, muons, photons, \tauhadvis, jets and $b$-jets in the final state.

Electron candidates consist of a track matched to a cluster of energy deposited in the electromagnetic calorimeter~\cite{EGAM-2021-01,EGAM-2021-02}.
Selection criteria for the track impact parameters are imposed to guarantee that the track originates from the primary vertex.
The transverse and longitudinal impact parameters, denoted by $d_0$ and  $z_0$, are required to satisfy $d_0 / \sigma(d_0) < 5$ and $|z_0 \sin\theta| < 5~\text{mm}$, where $\sigma(d_0)$ is the uncertainty in $d_0$ and $\theta$ is the track's polar angle.
The pseudorapidity of the calorimeter energy cluster must satisfy $|\eta_\text{cluster}| < 1.37$ or $1.52 < |\eta_\text{cluster}| < 2.47$.
The electron candidates are required to have $\pt > 15~\gev$ and satisfy the \textit{Loose}
ID selection criterion described in Ref.~\cite{EGAM-2018-01}.
The \textit{Loose} ID selection efficiency is greater than 88\% for all considered electron candidates.
A multivariate algorithm further suppresses non-isolated
electrons arising from $b$-hadron decays~\cite{HIGG-2017-02}.
The selection efficiency of the isolation criterion is greater than 70\% for electron candidates
with $\pt > 20~\gev$.

Photon candidates are reconstructed from energy clusters in the electromagnetic calorimeter similarly to electrons~\cite{EGAM-2021-01,EGAM-2021-02}.
Both converted (matched to a conversion vertex) and unconverted (not matched to an electron track or a conversion vertex) photons are considered.
All photon candidates are required to have $\pt > 15~\gev$.
The identification criterion used in the present study is the \textit{Tight} selection~\cite{EGAM-2018-01,EGAM-2021-01}.
Its selection efficiency is greater than 92\% (96\%) for unconverted (converted) photons with $\pt > 140~\gev$.
The \textit{Tight} isolation requirement defined in Ref.~\cite{PERF-2017-02} is also used.
The selection efficiency of this criterion is greater than 80\% for photons with $\pt > 140~\gev$.

Muon candidates consist of a track reconstructed in the inner detector matched to a track found in the muon spectrometer.
Information about the two tracks is combined to measure the muon momentum more precisely~\cite{PERF-2015-10}.
The resulting track is then required to satisfy the same criteria for the impact parameters as electron tracks.
Finally, muon candidates are selected if they have $\pt > 7~\gev$, $|\eta|<2.5$ and
satisfy the \textit{Loose} ID criterion described in Ref.~\cite{MUON-2018-03}.
The selection efficiency of the ID criterion is greater than 96\% for all considered muon candidates.
Muons are also required to satisfy isolation criteria similar to those for electrons.
The selection efficiency of the isolation requirement is greater than 80\% for muon candidates
with $\pt > 20~\gev$.

Jets are reconstructed from constituents built according to the particle-flow algorithm, which exploits both tracks
and calorimeter energy clusters~\cite{PERF-2015-09}.
The particle-flow objects are passed to the \antikt algorithm~\cite{Cacciari:2008gp,Fastjet} with a radius parameter of \(R=0.4\) to form jets via a four-momentum recombination scheme.
Jet energy is calibrated to the hadronic scale with the effect of \pileup removed~\cite{JETM-2018-05}.
In-situ jet calibration is based on measurements of $Z$\,+\,jets, $\gamma$\,+\,jets and multijet events.
Jets are required to have $\pt > 20~\gev$ and rapidity $|y| < 4.5$.

A multivariate $b$-tagging algorithm named DL1r~\cite{FTAG-2019-07} is used to tag
jets from $b$-quarks.
It exploits information about the impact parameters of tracks and properties of vertices in the jets.
The algorithm tags jets with $\pt > 20~\gev$ and $|\eta| < 2.5$.
A fixed 85\%-efficiency working point is used, where the efficiency is evaluated in \ttbar events from MC simulation.
The flavour-tagging efficiency for jets from $b$, $c$, and lighter quarks is calibrated as described in Refs.~\cite{FTAG-2018-01,FTAG-2019-02,FTAG-2020-08}.

Another multivariate algorithm, the jet-vertex tagger (JVT)~\cite{PERF-2014-03}, is used to veto jets
with $\pt<60~\gev$ and $|\eta|<2.5$ if their constituent tracks are not consistent with the primary event vertex,
as indicated by a low JVT output score.
Similarly, a forward jet-vertex tagging (fJVT) algorithm~\cite{PERF-2016-06-witherratum} vetoes such jets
with $\pt<60~\gev$ and $|\eta|>2.5$.

The \tauhadvis candidates~\cite{ATLAS-CONF-2017-029,ATL-PHYS-PUB-2015-045,PERF-2013-06} are seeded by jets formed
using the \antikt algorithm with a radius parameter of \(R=0.4\).
Three-dimensional clusters of calorimeter energy deposits called topo-clusters~\cite{PERF-2014-07}, calibrated using a local
hadronic calibration (LC)~\cite{ATL-LARG-PUB-2009-001-2}, serve as inputs to the jet algorithm.
The transverse momentum of a \tauhad cannot be fully reconstructed because of an undetected neutrino.
The energy scale of the visible decay products (\tauhadvis) is measured using \Ztt events with one \tauhadvis and one \tl-lepton decaying into a muon and neutrinos.
Hadronic decay modes belonging to five categories are reconstructed~\cite{PERF-2014-06}: decays with one matched track and either zero, one or more neutral pions,
and decays with three matched tracks and either zero or more neutral pions.
However, the present study groups the five categories into two to reduce the statistical uncertainties.
The two categories consist of decays with either one or three matched tracks, referred to as 1-prong (1p) and 3-prong (3p) decays respectively.
Only \tauhadvis with reconstructed electric charge $|q| = 1$ are selected.
Each \tauhadvis is required to have $\pt > 20~\gev$, $|\eta|<1.37$ or $1.52<|\eta|<2.5$,
and a score greater than 0.01 from an ID algorithm based on recurrent neural networks (RNN)~\cite{ATL-PHYS-PUB-2019-033}.
In the present analysis, regions of phase space are often split into two parts according to the \tauhadvis ID score.
The score value corresponding to the \textit{Loose} ID working point is used for this purpose.
Its efficiency is 85\% (75\%) for 1-prong (3-prong) \tauhadvis candidates regardless of their \pt.
Its rejection power is 21 (90) for 1p (3p) \tauhadvis candidates in the dijet MC simulation.

Since particles are reconstructed by independent algorithms for each type of object, the same
detector energy depositions are often reconstructed as more than one type of object.
For example, a \tauhadvis is almost always reconstructed as a jet.
It is necessary to remove this ambiguity and keep only one of the overlapping objects.
The procedure for removing the redundant reconstructed objects is described in Ref.~\cite{HIGG-2019-09}.
A \tauhadvis object is removed if an electron or muon is reconstructed within an angular distance of
$\Delta R < 0.2$.
A jet is removed if there is a \tauhadvis within $\Delta R < 0.2$.

The missing transverse momentum \ptmis is defined as the negative vectorial sum of transverse momenta of
electrons, muons, photons, \tauhadvis, and jets.
Soft particle tracks not matched to any object are taken into account via a separate
'soft term'~\cite{JETM-2020-03}.
The magnitude of the \ptmis is denoted by \met.

When estimating the true \tl-lepton, electron or muon contribution using the MC samples, each \tauhadvis candidate
is required to geometrically match the generator-level \tl-lepton, electron, or muon.
In this situation, jets from the hadronization of quarks or gluons misidentified as \tauhadvis are not
considered.
The contamination of the fake-enriched samples by \tl-leptons, electrons or muons is called real
\tl-lepton contamination in the following.


%

%

\section{The fake-factor method}
\label{sec:ff}

Assuming a SR with exactly one \tauhadvis candidate in the final state
and assuming the candidate is a fake-\tauhadvis,
the number of events in the ID subregion of the SR, $N^\mathrm{fake}_\mathrm{ID}$,
can be estimated as the number of events in the antiID subregion, $N^\mathrm{fake}_\mathrm{antiID}$, times a constant called a fake factor, \ff:
\begin{equation}
\label{eq:ffm}
N^\mathrm{fake}_\mathrm{ID} = N^\mathrm{fake}_\mathrm{antiID} \times \ff.
\end{equation}
The \ff value can be estimated using an additional fake-enriched DR:
\begin{equation}
\label{eq:ff}
\ff = \frac{ N^\mathrm{fake}_\mathrm{ID\ DR} }{ N^\mathrm{fake}_\mathrm{antiID\ DR} }.
\end{equation}
The DR is split into two subregions, defined by the same selection criteria except for the \tauhadvis ID; $N^\mathrm{fake}_\mathrm{ID\ DR}$ ($N^\mathrm{fake}_\mathrm{antiID\ DR}$) stands for the number of DR events in which the \tauhadvis passes (fails) the ID selection.
The DR selection criteria are typically very similar to the SR ones.
Often, just one criterion is changed, e.g.\ the requirement on the \tauhadvis and a light lepton charges product in analyses targeting final states with two \tl-leptons.

\subsection{Details of the fake-factor method}
\label{sec:ff_real_tau}

The fake-enriched regions, the antiID subregion of the SR and the DR, also select a certain number of events where
the reconstructed \tauhadvis candidate is not a jet originating from a QCD process.
However, the FF method is defined just for fake-\tauhadvis events, and
any contamination by real-\tl-leptons must be removed.
The real-\tl-lepton contamination is estimated with MC simulation,
by geometrically matching \tauhadvis candidates to generator-level objects as described in Section~\ref{sec:eventreco},
and then subtracted from the data event yield, so that Eqs.~(\ref{eq:ffm}) and~(\ref{eq:ff}) become:
\begin{eqnarray} \nonumber
N^\mathrm{fake}_\mathrm{ID} &=& \left(N^\mathrm{data}_\mathrm{antiID} - N^\mathrm{real}_\mathrm{antiID}\right) \times \ff, \label{eq:ff_1tau} \\ \nonumber
\ff &=& \frac{ N^\mathrm{data}_\mathrm{ID\ DR} - N^\mathrm{real}_\mathrm{ID\ DR} }{ N^\mathrm{data}_\mathrm{antiID\ DR} - N^\mathrm{real}_\mathrm{antiID\ DR} } \label{eq:ff_formula},
\end{eqnarray}
where the superscripts \enquote{data} and \enquote{real} label the data event yields and MC-estimated real-\tl-lepton event yields, respectively.

Another feature of the method is that the \ff is not a constant, but depends on the properties of the fakes.
The present study takes into account its dependence on \pt and the number of matched tracks (\enquote{prongness}).
A parameterization as a function of additional variables, such as $\eta$, was considered,
but discarded because there were too few events.
Section~\ref{sec:validation} shows that the fakes' $\eta$ spectra are well modelled by the UFF method's predictions.
The two-dimensional phase space is split into bins of \pt and prongness in which the FF method is used as described above.
The final fake-background estimate for the SR is a sum of fake-background estimates coming
from the individual phase-space bins.

The above fake-background estimation algorithm is thus easy to generalize for the case of multi-bin SRs.
In each SR bin, the estimation is performed independently and as in the single-bin SR case.

\subsection{Composition of fake-\tauhadvis samples}
\label{sec:ff_composition}

Ideally, the DR where the FF is determined should be defined to be as similar to the SR as possible
in order to obtain fake-\tauhadvis samples with the same composition.
In general, fake \tauhadvis can arise from several types of jets.
Four main sources of fakes are considered: $b$-quarks (\bfake); light quarks (\qfake);
gluons, (\gfake); and jets arising from \pileup (\pfake).
Fakes from these sources have different properties,
and therefore these four types of jets have different probabilities of being misidentified as a \tauhadvis.
The FF varies among the types of fakes and, in general, it is not possible to measure the \ff
for each type of fake separately because every region of phase space contains a mixture of the different types.
An ideal DR for the \ff measurement would contain a sample of fakes with the same \qgbp composition
as in the SR, and fake-composition differences between the DR and SR are a source of systematic uncertainty.

\subsection{Fake-factor method with two \tauhadvis in the final state}
\label{sec:ff_2taus}

The FF method can be generalized for final states containing two \tauhadvis.
Often, there are two major sources of fakes, one producing a final state with
a real \tauhadvis and a fake, and the other producing two fakes in the final state.
In these cases, the dominant fake processes are typically $\wtaunu$\,+\,jets and multijet production,
and the FF method formula is:
\begin{eqnarray}
\label{eq:ff_2tau}
N^\mathrm{fake}_\mathrm{2\;ID} & = & \left(N^\mathrm{data}_\mathrm{1\;ID,\;1\;antiID} - N^\mathrm{real}_\mathrm{1\;ID,\;1\;antiID}\right) \times \ff \nonumber \\
& & - \left(N^\mathrm{data}_\mathrm{2\;antiID} - N^\mathrm{real}_\mathrm{2\;antiID}\right) \times \ff'_1 \times \ff'_2,
\end{eqnarray}

where the event yields $N$ have subscripts to denote ID and antiID \tauhadvis candidates,
and superscripts to denote the data, real \tauhadvis, and fake samples.
In general, the fake-factor sets $\ff$ and $\ff'$ are different from each other.
$\ff'_1$ and $\ff'_2$ are taken from the same fake-factor set but they can have different values,
depending on the \pt and prongness of the fakes.
The subtraction of the two-fake yield in the second term of Eq.~(\ref{eq:ff_2tau}) is needed to
cancel out the double counting of events with two fakes in the first term.
The double counting occurs for combinatorial reasons:\
events where one \tauhadvis candidate passes the ID selection and the second one fails
are as likely as events where the two \tauhadvis candidates are swapped.
Since both groups provide the full estimate of the two-fake event yield,
double counting occurs.


%

%

\section{The Universal Fake Factor method}
\label{sec:uff}

The UFF method provides a prescription for estimating the FFs suitable for the SR of an analysis.
The method is a general solution to the major drawback of the FF method discussed in Section~\ref{sec:ff_composition}.
The need to find a DR with the same \qgbp composition as the SR is a serious limitation of the FF method.
Rather than finding a suitable DR for each SR, the UFF method determines the fractions of the four main fake
categories composing the antiID subregion of the SR, and uses them to evaluate the appropriate \ff.
Ideally, the categories would be \qfake, \gfake, \bfake, and \pfake fakes.
However, it is impossible to find regions containing only one type of fake in order to measure
the \qgbp-fake components for the UFF method.
Instead, regions enriched in each type of fake are used.

\subsection{Fake-enriched regions}
\label{sec:uff_regions}

Four selections are used to separate regions of phase
space enriched in each category of fake background.
The \Zmm region is enriched in \qfake fakes, while the \ttbar region is enriched in \bfake fakes.
Two multijet regions, called \mjh and \mjl, are enriched in \gfake and \pfake fakes, respectively.
Each region's definition requires the presence of exactly one \tauhadvis candidate.
The \tauhadvis must pass the \textit{Tight} BDT-based electron veto, eBDT~\cite{HIGG-2019-09}, to suppress electrons misidentified as \tauhadvis.
Its efficiency is 75\% for \tauhadvis candidates passing the ID requirement.
The veto is applied only to 1p \tauhadvis candidates.

The \Zmm region consists of events with exactly two opposite-charge muons.
The muons must meet the \textit{Tight} ID criterion~\cite{PERF-2015-10} and have $\pt > 15~\gev$.
The mass of the dimuon system, \mumum, is required to be within the range $76{-}106~\gev$.
The events are triggered by a logical OR of two single-muon trigger items~\cite{TRIG-2018-01,TRIG-2019-03}.
One of them requires a muon with an online-reconstructed $\pt > 50~\gev$, while the other one has
a lower \pt threshold of $26~\gev$ ($20~\gev$ in 2015 data) but imposes an additional isolation requirement
on the muon candidate.
The leading muon is further required to have an offline $\pt > 28~\gev$.
Events with an electron, photon or $b$-jet are vetoed.
The seed jet of the \tauhadvis candidate must have a \jvt score $> 0.8$.

The \ttbar region is defined by the presence of one electron, one muon and at least one $b$-jet.
The muon has to satisfy the \textit{Tight} ID requirement and have $\pt > 10~\gev$.
The electron is required to meet the \textit{Tight} LLH ID criterion and have $\pt > 27~\gev$.
The two leptons must have opposite charges.
The events are triggered by a logical OR of three single-electron trigger items~\cite{TRIG-2018-05,TRIG-2019-03}.
Their online \pt thresholds are 26, 60 and 140~\gev (24, 60 and 120~\gev in 2015 data), and they impose
\textit{Tight}, \textit{Medium} and \textit{Loose} (\textit{Medium}, \textit{Medium} and \textit{Loose}
in 2015 data) online ID requirements on the electron candidate,
as well as an isolation requirement on the L1 objects.
A photon veto is applied.
The \tauhadvis candidate's seed jet must have a \jvt score $> 0.8$.

Two multijet regions are defined by the presence of at least two jets that are back-to-back in the transverse plane.
The events are triggered by a single-jet trigger~\cite{TRIG-2012-01} with an online \pt threshold of $420~\gev$ ($360~\gev$ in 2015 data).
Both jets must have $|\eta| < 3$, and the leading jet's \pt is required to exceed $450~\gev$ ($390~\gev$ in 2015 data).
The back-to-back topology is ensured by requiring the two leading jets to have an azimuthal separation of at least 2.7.
The \tauhadvis candidate's \pt must be less than one-fourth of the two leading jets' average \pt
to enrich the sample in \gfake fakes.
For a similar reason, a third-jet veto is applied.
The events are required to have either exactly two jets or a third jet with \pt lower than the \tauhadvis candidate's \pt.
Events with an electron, muon, photon or $b$-jet are vetoed.
One multijet region, called \mjh, is defined by imposing a requirement of $\jvt > 0.8$ on the \tauhadvis candidate's seed jet.
The other multijet region, called \mjl, requires the \tauhadvis candidate's seed jet to have $\jvt < 0.8$.

The purity of the target process in the \Zmm, \ttbar, \mjh and \mjl regions is estimated using the Powheg MC sample set.
In the \Zmm region, \Zmm production accounts for more than 95\% of the events,
while multijet production contributes more than 95\% of the events in the \mjh and \mjl regions.
In the \ttbar region, \ttbar production constitutes over 90\% of the events.
The contamination from real-\tl-leptons, estimated with the Default MC sample set,
is below 10\% in the \Zmm, \mjh and \mjl regions, and below 25\% in the \ttbar region.

Two regions are defined to estimate the systematic uncertainty of the UFF method, and to test the method.
The \wmunu region is defined by the presence of exactly one muon and one \tauhadvis, with the same electric charge.
The \tauhadvis candidate must satisfy the \textit{Tight} eBDT criterion and be separated from the muon by $\mutaudr >0.4$ in the $\eta{-}\phi$ plane.
The muon is required to have $\pt > 27.3~\gev$ and satisfy the \textit{Medium} ID criterion.
The trigger strategy is the same as in the \Zmm region.
Events with an electron or photon are vetoed.
The \muthreej events form a subregion of the \wmunu region.
In addition to the \wmunu requirements, the \muthreej region definition requires the presence of at least three jets, and at least one of them must be $b$-tagged.

One more region, called \gl, is defined to estimate the systematic uncertainty of the UFF method and consists of events with a photon and a \tauhadvis candidate.
A single-photon trigger with an online \pt threshold of $140~\gev$ and the \textit{Loose} ID requirement~\cite{EGAM-2018-01} is used.
The photon is required to have $\pt > 145~\gev$.
The \tauhadvis candidate must satisfy the \textit{Tight} eBDT criterion, and its seed jet must have $\jvt < 0.8$.
The distance \phtaudr must exceed 0.2.
Events with an electron or muon are vetoed.

Table~\ref{tab:regions} summarizes the selection criteria defining the \Zmm, \ttbar, \mjh, and \mjl regions.
Figure~\ref{fig:uff_composition} shows the \qgbp fake composition of the antiID \Zmm, \ttbar, \mjh, \mjl, \wmunu, and \muthreej subregions,
with an inclusive selection in \pt and prongness.
The composition is determined with the Powheg set of MC samples.
Regions \Zmm, \ttbar, \mjh, and \mjl are each enriched in the type of fake they target.
The \gl region is not shown in the figure because it has a limited usage in the pile-up related systematic uncertainty estimation.
\begin{figure}
\centering
\includegraphics[width=0.45\textwidth]{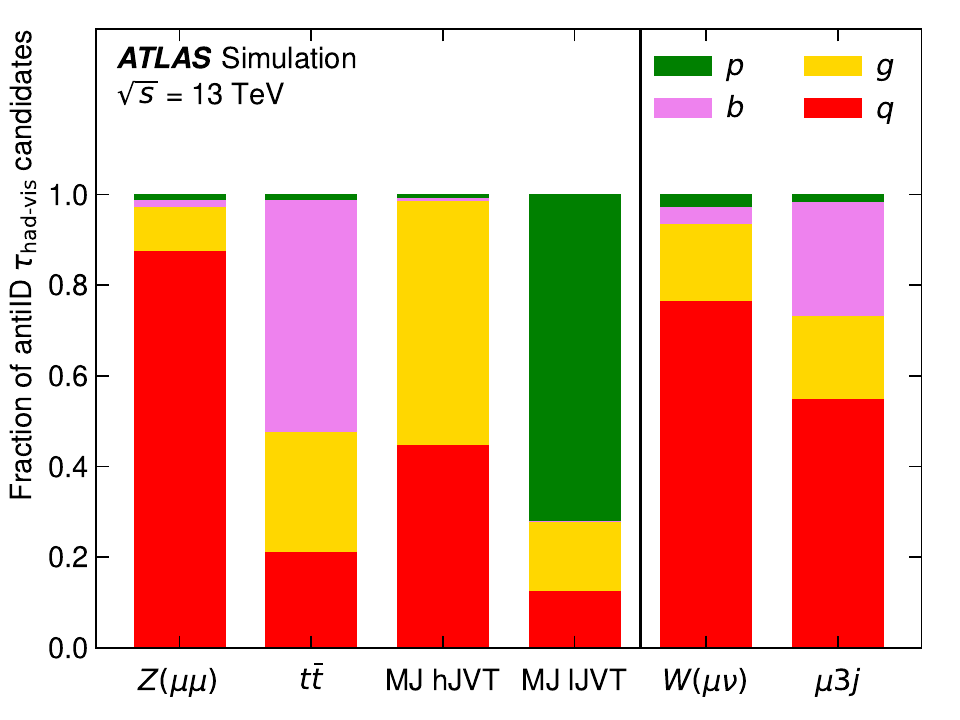}
\caption{
The \qgbp fake composition of the antiID \Zmm, \ttbar, \mjh, \mjl, \wmunu, and \muthreej subregions,
with an inclusive selection in \pt and prongness.
}
\label{fig:uff_composition}
\end{figure}

\renewcommand{\arraystretch}{1.2} %
\begin{table}[htbp]
\captionsetup{aboveskip=10pt, belowskip=10pt}
\centering
\caption{Selection criteria defining the \Zmm, \ttbar, \mjh, and \mjl regions.
The symbols $\mu$, $e$, $j$, $\gamma$ stand for a muon, an electron, a jet, and a photon, respectively.
Their numerical subscripts refer to the ordering of that type of object in \pt.
The symbol $q$ stands for the electric charge.
The azimuthal separation of the two leading jets is denoted by \jjdphi.
}
\begin{tabular}{c|c|c|c}
\toprule
\Zmm & \ttbar & \multicolumn{2}{c}{MJ} \\
\hline
$N_\mu = 2$ & $N_\mu = 1$ & \multicolumn{2}{c}{$N_\mu = 0$} \\
$N_e = 0$ & $N_e = 1$ & \multicolumn{2}{c}{$N_e = 0$} \\
$N_{b\text{-jet}} = 0$ & $N_{b\text{-jet}} \geq 1$ & \multicolumn{2}{c}{$N_{b\text{-jet}} = 0$} \\
$N_{\gamma} = 0$ & $N_{\gamma} = 0$ & \multicolumn{2}{c}{$N_{\gamma} = 0$} \\
Single-$\mu$ trigger & Single-$e$ trigger & \multicolumn{2}{c}{Single-jet trigger} \\
\multirow{2}{*}{\textit{Tight} $\mu$ ID} & \textit{Tight} LLH $e$ ID & \multicolumn{2}{c}{} \\
& \textit{Tight} $\mu$ ID & \multicolumn{2}{c}{} \\
$\pt^{\mu_{1}} > 28~\gev$ & $\pt^e > 27~\gev$ & \multicolumn{2}{c}{$\pt^{j_1} > 450~\gev$} \\
$\pt^{\mu_{2}} > 15~\gev$ & $\pt^{\mu} > 10~\gev$ & \multicolumn{2}{c}{} \\
& & \multicolumn{2}{c}{$|\eta_{j_1}| < 3$} \\
& & \multicolumn{2}{c}{$|\eta_{j_2}| < 3$} \\
$q_{\mu_{1}} q_{\mu_{2}} = -1$ & $q_{e} q_{\mu} = -1$ & \multicolumn{2}{c}{} \\
$\mumum \in (76, 106)~\gev$ & & \multicolumn{2}{c}{$\jjdphi > 2.7$} \\
& & \multicolumn{2}{c}{$\pttau < \left( \pt^{j_1} + \pt^{j_2} \right) \Big/ 8$} \\
& & \multicolumn{2}{c}{$N_j = 2 \ \mathrm{OR} \ \pt^{j_3} < \pttau$} \\
\textit{Tight} \tauhadvis eBDT & \textit{Tight} \tauhadvis eBDT & \multicolumn{2}{c}{\textit{Tight} \tauhadvis eBDT} \\
\multirow{2}{*}{$\text{JVT}_{\tauhadvis} > 0.8$} & \multirow{2}{*}{$\text{JVT}_{\tauhadvis} > 0.8$} & \mjh & \mjl \\
& & $\text{JVT}_{\tauhadvis} > 0.8$ & $\text{JVT}_{\tauhadvis} < 0.8$ \\
\bottomrule
\end{tabular}
\label{tab:regions}
\end{table}

\subsection{Evaluation of combined fake factors}
\label{sec:uff_idea}

The four fake-enriched regions, \Zmm, \ttbar, \mjh, and \mjl, are split into \pt and prongness bins.
The same is done for the antiID subregion of the SR of the analysis employing the UFF method.
In each bin, the antiID subregion of the SR consists of four subregions.
The fake population in each of these subregions has a composition similar to that of a mixture
of the fake populations in the antiID subregions of the \Zmm, \ttbar, \mjh, and \mjl regions.
The UFF method estimates this mixture of subregions contributing to the antiID subregion of the SR.
This is done using a template fit of the ATLAS q/g tagger score~\cite{JETM-2020-02} spectra,
which will be discussed more in detail in Section~\ref{sec:uff_fit}.
The combined FF suitable for the SR in question is then calculated as a weighted
average of FFs measured in the \Zmm, \ttbar, \mjh, and \mjl regions:
\begin{equation}
\label{eq:uff}
\ffuff = \muz \times \ffz + \mutt \times \fftt + \mumjh \times \ffmjh + \mumjl \times \ffmjl,
\end{equation}
where the individual FFs and their weights in the combination are denoted by
$\ff_X$ and $\mu_X$, respectively, and $X = \Zmm, \ttbar, \mjh, \mjl$.
The FFs are shown in Figure~\ref{fig:uff_ff}.
They decrease as a function of the \tauhadvis candidate's \pt because the fake rejection by the RNN algorithm improves
with increasing \pt~\cite{ATL-PHYS-PUB-2019-033}.
The FFs for the 1-prong fakes are higher than for the 3-prong fakes
because the RNN's rejection power is higher for the latter.
FFs determined in the \mjl region tend to be lower than others,
due to \pileup track rejection from the track matching and classification
applied in the \tauhadvis reconstruction.
\mjh FFs are lower than \Zmm FFs because there is a higher admixture of \gfake fakes in the \mjh region,
and the \gfake fakes have lower FFs than the \qfake fakes, as discussed in Section~\ref{sec:uff_unc}.
\begin{figure}
\centering
\subfloat[]{
\includegraphics[width=0.45\textwidth]{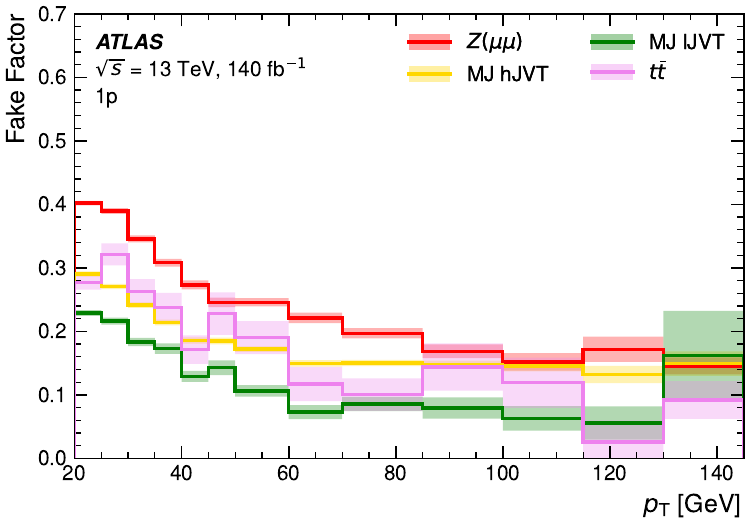}
\label{fig:uff_ff_1p}
}\quad
\subfloat[]{
\includegraphics[width=0.45\textwidth]{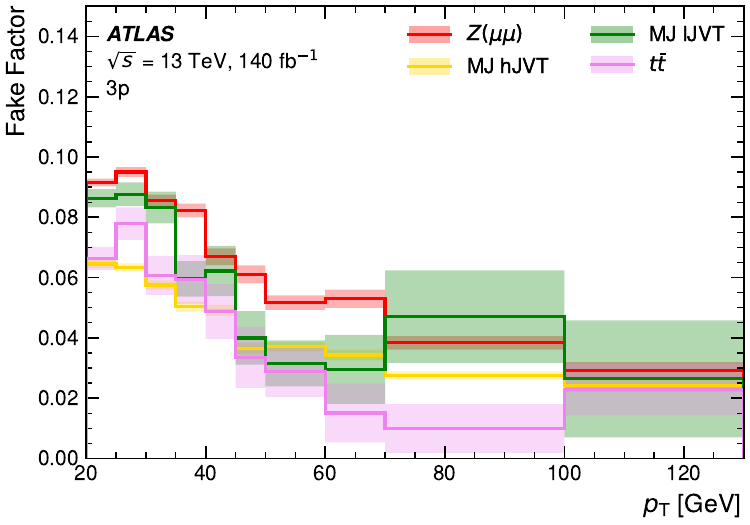}
\label{fig:uff_ff_3p}
}
\caption{
FFs estimated in the \Zmm, \ttbar, \mjh and \mjl regions
for (a) 1-prong (1p) fakes and (b) 3-prong (3p) fakes as a function of the fake's \pt.
The coloured bands display the statistical uncertainty of the FFs.
The highest \pt bin contains events with fake-\tauhadvis with \pt of up to 300~\gev.
}
\label{fig:uff_ff}
\end{figure}

It should be noted that the FFs from the \Zmm, \ttbar, \mjh and \mjl regions
can be expressed as linear combinations of the \qgbp FFs.
The UFF method assumes that the same is true for the SR FFs.
Therefore, there is no approximation in Eq.~(\ref{eq:uff}) when using the \Zmm, \ttbar, \mjh and \mjl FFs
instead of the \qgbp ones.

\subsection{Template fit}
\label{sec:uff_fit}

In each \pt and prongness bin and each antiID subregion, a normalized template of the ATLAS q/g tagger
score is built.
The q/g tagger provides a multivariate-analysis score based on jet variables that are sensitive to the differences
in radiation patterns, fragmentation, and hadronization between quark-initiated and gluon-initiated jets.
Beside its sensitivity to \qfake and \gfake fakes,
it also has a certain sensitivity to the properties of the \bfake and \pfake fakes.
The q/g tagger score is evaluated on a particle-flow jet (see Section~\ref{sec:eventreco}),
closest to the \tauhadvis candidate in the $\eta{-}\phi$ plane in each event.
If no particle-flow jet is found within a $\Delta R = 0.4$ distance from the \tauhadvis candidate,
the candidate is added to the lowest q/g tagger score bin.
Each template is built using data, as all the regions are fake-enriched.
The residual real-\tl-lepton contamination is subtracted from the \Zmm, \ttbar, \mjh and \mjl templates via MC simulation.
Then, the four templates are normalized to the event yield in the antiID subregion of the SR minus the real-\tl-lepton
contamination in that region.
The Default set of MC samples is used to estimate the real-\tl-lepton contamination.
As an example, Figure~\ref{fig:uff_templates} shows the four templates
in the 1-prong and $\pt \in (85, 100)~\gev$ bin.
For display purposes, the templates are normalized to unity.
The sum of the \Zmm, \ttbar, \mjh, and \mjl templates plus the histogram of the real-\tl-lepton contamination in the antiID subregion of the SR is then fitted
to the data in the antiID subregion of the SR using the maximum-likelihood method~\cite{Baak:2014wma,Cranmer:2012sba,roostats,RooFit,James:1994vla}.
The fit is performed independently in each \pt and prongness bin.
The normalization factors of the \Zmm, \ttbar, \mjh, and \mjl templates are free parameters of the fit and their post-fit values are used as estimates of the
combination coefficients $\mu_X$ in Eq.~(\ref{eq:uff}).
No normalization factor is assigned to the real-\tl-lepton contamination in the antiID subregion of the SR.
The uncertainty due to the limited size of the samples for the \Zmm, \ttbar, \mjh and \mjl templates
and the real-\tl-lepton contamination in the antiID subregion of the SR is taken into account in the fit.
It is parameterized by nuisance parameters with Poisson constraints assigned to each bin of the q/g tagger
score variable.
An important feature of the UFF method is that the normalization factors are not constrained to be positive.
This allows the method to provide an extrapolation in situations where the antiID subregion of the SR is more enriched in one type of
fake than any antiID subregion of \Zmm, \ttbar, \mjh, and \mjl events.
In these cases, the normalization factor corresponding to the region that is the most enriched in the given
fake type can be larger than one.
Consequently, at least one other normalization factor is usually negative so that
the sum of the four normalization factors is close to one.
Although the fit does not constrain the sum to one, its value is approximately one
due to the absence of nuisance parameters substantially affecting the template normalization.
As an example, Figure~\ref{fig:uff_fit_result} shows the results of the template fit in the 1-prong and
$\pt \in (85, 100)~\gev$ bin of the \wmunu region.
In the figure, the q/g tagger score templates of the \Zmm, \ttbar, \mjh and \mjl antiID subregions are shown,
and they are scaled by their corresponding post-fit normalization factors.
The contribution from real-\tl-lepton background is also shown in the figure.
Figure~\ref{fig:uff_fit_fractions} shows the breakdown plot of the post-fit \Zmm, \ttbar, \mjh, and \mjl templates
whose sum is normalized to unity in each q/g tagger score bin.
\begin{figure}
\centering
\includegraphics[width=0.45\textwidth]{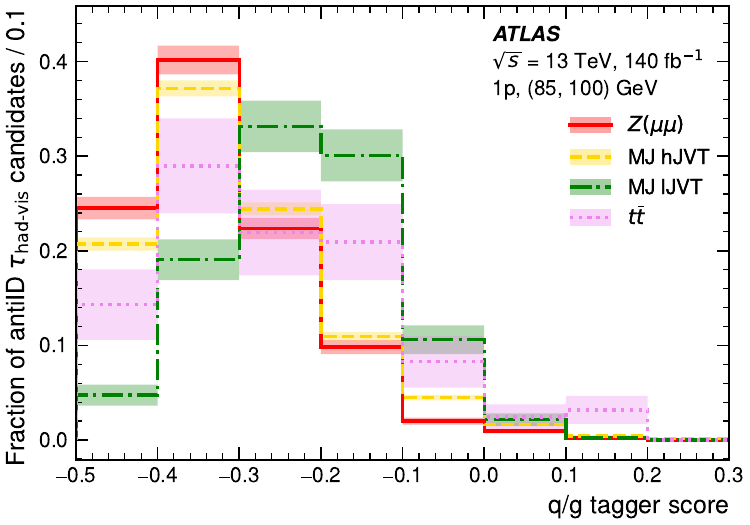}
\caption{
Templates of the q/g tagger score distributions in the 1-prong (1p) and $\pt \in (85, 100)~\gev$ bin.
They are determined in the \Zmm, \ttbar, \mjh and \mjl antiID subregions.
For display purpose, the templates are normalized to unity.
The bands display the statistical uncertainty of the four templates.
}
\label{fig:uff_templates}
\end{figure}

\begin{figure}
\centering
\subfloat[]{
\includegraphics[width=0.45\textwidth]{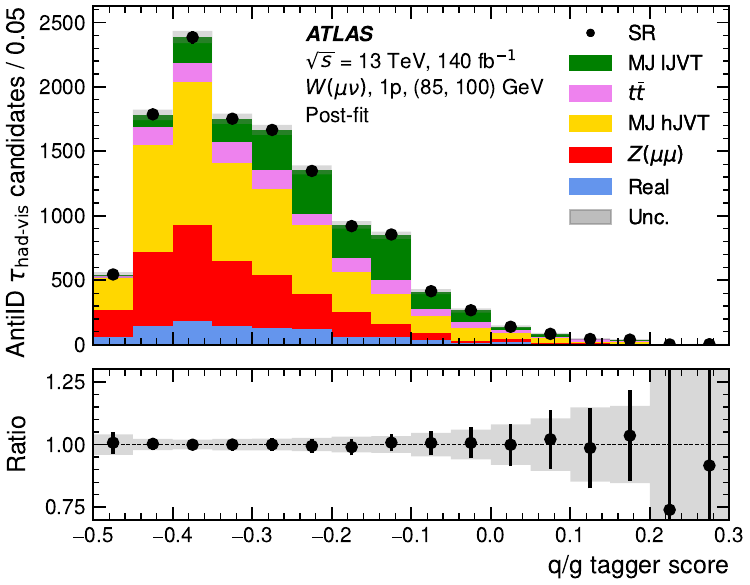}
\label{fig:uff_fit_result}
}\quad
\subfloat[]{
\includegraphics[width=0.45\textwidth]{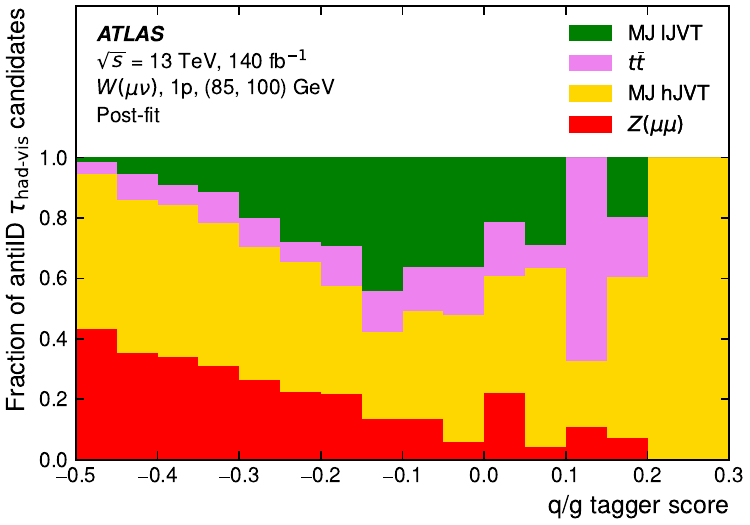}
\label{fig:uff_fit_fractions}
}
\caption{
(a) Results of the template fit in the 1-prong and $\pt \in (85, 100)~\gev$ bin of the \wmunu region.
The black points represent the q/g tagger score distribution from the \wmunu antiID subregion.
The coloured histogram areas show the templates of the \Zmm, \ttbar, \mjh and \mjl antiID subregions
and the real-\tl-lepton background in the antiID subregion of the SR.
The four normalized templates are scaled by their corresponding post-fit normalization factors.
The grey band displays the statistical uncertainty of the sum of the four templates and the real-\tl-lepton background.
The post-fit normalization factors are $\mu_{\Zmm} = 0.28 \pm 0.16$, $\mu_{\ttbar} = 0.10 \pm 0.09$, $\mu_{\mjh} = 0.45 \pm 0.19$, $\mu_{\mjl} = 0.18 \pm 0.06$.
(b) Breakdown plot of the post-fit \Zmm, \ttbar, \mjh, and \mjl templates.
Their sum is normalized to unity in each q/g tagger score bin so that each template contribution
had the meaning of a fraction of the antiID \tauhadvis candidates in the bin.
}
\label{fig:uff_fit}
\end{figure}

\subsection{Universal Fake Factor method uncertainties}
\label{sec:uff_unc}

The \Zmm, \ttbar, \mjh and \mjl FFs and the coefficients $\mu_X$ from Eq.~(\ref{eq:uff}) are estimated
using samples of finite size.
The statistical uncertainties of the FFs are propagated to \ffuff assuming that
these uncertainties are uncorrelated.
On the other hand, the uncertainties of the coefficients $\mu_X$ are correlated,
and the Hessian covariance matrix estimate, $V$, is taken into account in the uncertainty propagation.
The total statistical uncertainty of \ffuff is estimated as the square root of the following expression:
\begin{equation*}
\label{eq:uff_stat}
\sigma^2_\mathrm{UFF,\ stat.} = \sum_{X} \mu_X^2 \sigma_{\ff_X}^2 + \sum_{X, Y} \ff_X \ff_Y V_{XY},
\end{equation*}
where $X$ and $Y$ run over the \Zmm, \ttbar, \mjh and \mjl regions.

The UFF method assumes that the FFs for \qgbp fakes are the same in all regions of phase space.
For \qgb fakes, this assumption is tested using MC simulation by comparing the FFs for a given fake type in five different regions.
The regions are carefully chosen to represent common fake-production processes and kinematic regimes.
Three of these five regions are \Zmm, \ttbar and \mjh
\footnote{\mjl is not in this list, as it targets \pfake fakes, for which a dedicated systematic uncertainty is defined.}
defined in Section~\ref{sec:uff_regions}.
The fourth region, \wmunu, targets \qfake fake production in association with a $W$ boson.
The fifth region, \muthreej, targets semileptonic \ttbar events,
containing \qfake fakes from the decay of a colourless particle, the $W$ boson.

In each region, the FFs for the \qfake, \gfake and \bfake fakes are evaluated
using the Powheg set of MC samples.
To reduce the effective statistical uncertainty, MC samples corresponding to just the dominant
process in each region are used, i.e. \zjets MC is used for the \Zmm region, \ttbar MC for the \ttbar region, etc.
Figure~\ref{fig:uff_ff_syst_q} shows the FFs for the \qfake fakes in the five regions.
They are rebinned and smoothed as a function of \pt to reduce the statistical fluctuations.
Bins are combined iteratively from the highest \pt bin with non-zero content until the FF is non-zero with a statistical significance of $1.5\sigma$,
or the lowest \pt bin is reached.
As the next step, FF values in the wide bins are copied to finer bins in which the nominal UFF prediction is evaluated.
For the smoothing, each FF set is further copied to an equidistant binning with a bin width of 1.
Then, it is smoothed using a Gaussian kernel with a width of 2.
A systematic uncertainty is assigned to the \Zmm FFs shown in Figure~\ref{fig:uff_ff}
by using the differences between the \Zmm MC FFs and the others to form the
symmetrized uncertainty envelope shown in Figure~\ref{fig:uff_ff_syst_q}.
In each \pt and prongness bin, the largest deviation of a FF from the \Zmm MC FF defines the envelope.
Subsequently, this is divided by the \Zmm MC FF to get the relative systematic uncertainty.
A similar procedure assesses the uncertainty of the \ttbar and \mjh FFs.
The envelope around the \ttbar (\mjh) MC FFs in Figure~\ref{fig:uff_ff_syst_b} (Figure~\ref{fig:uff_ff_syst_g})
divided by the \ttbar MC \bfake FFs (\mjh MC \gfake FFs) is used as the relative
symmetric systematic uncertainty of the \ttbar (\mjh) FFs.
When estimating the uncertainty of the \ttbar FFs, the \wmunu FFs are removed from the envelope
because the \wmunu sample is very small, implying no need to include it.
\begin{figure}
\centering
\subfloat[]{
\includegraphics[width=0.45\textwidth]{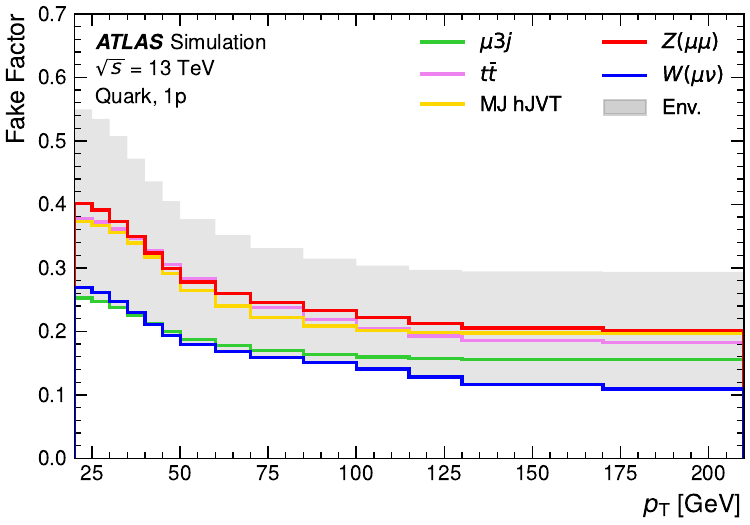}
\label{fig:uff_ff_syst_q_1p}
}\quad
\subfloat[]{
\includegraphics[width=0.45\textwidth]{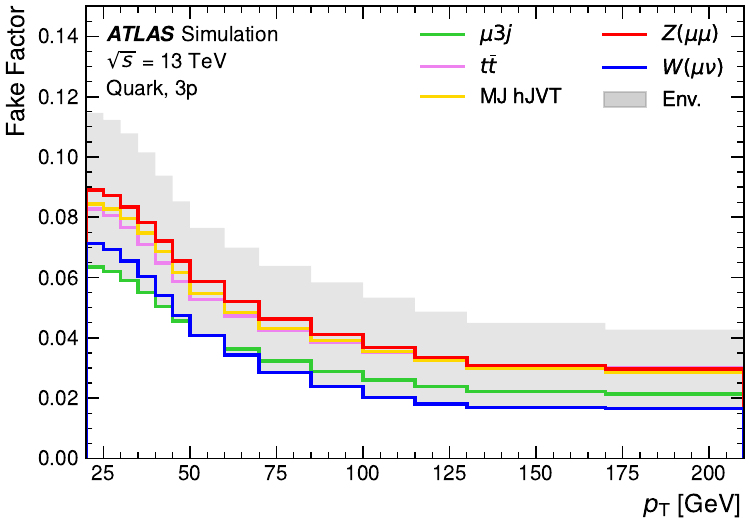}
\label{fig:uff_ff_syst_q_3p}
}
\caption{FFs for the (a) 1-prong (1p) \qfake fakes and (b) 3-prong (3p) \qfake fakes as a function of the fake-\tauhadvis \pt in five regions of phase space, evaluated using the MC simulation and smoothed.
The grey band displays the symmetrized envelope around the \Zmm FFs.
The envelope is used to estimate the systematic uncertainty of the \Zmm FFs.
The highest \pt bin contains events with fake-\tauhadvis with \pt of up to 300~\gev.
}
\label{fig:uff_ff_syst_q}
\end{figure}

\begin{figure}
\centering
\subfloat[]{
\includegraphics[width=0.45\textwidth]{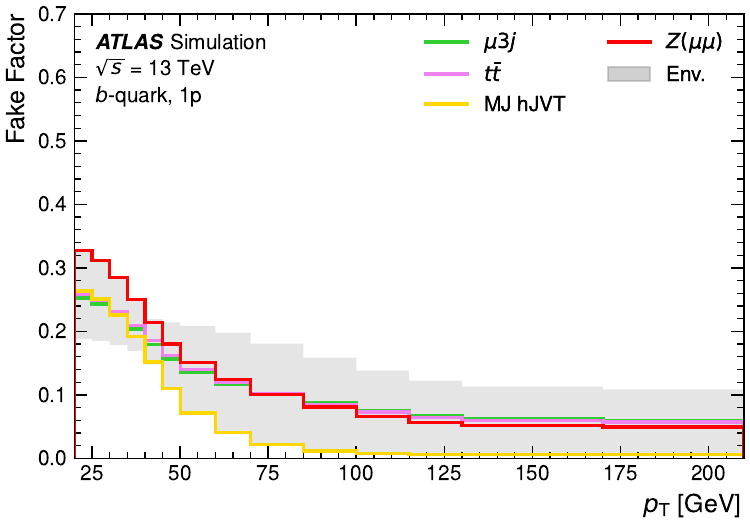}
\label{fig:uff_ff_syst_b_1p}
}\quad
\subfloat[]{
\includegraphics[width=0.45\textwidth]{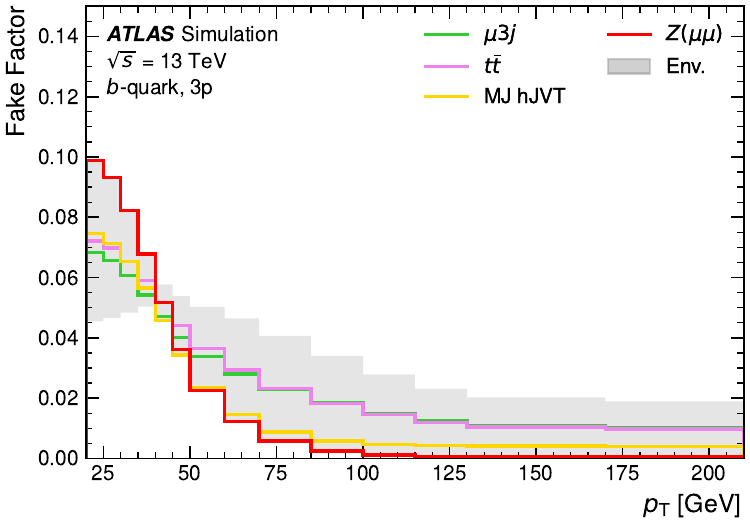}
\label{fig:uff_ff_syst_b_3p}
}
\caption{FFs for the (a) 1-prong (1p) \bfake fakes and (b) 3-prong (3p) \bfake fakes as a function of the fake-\tauhadvis \pt  in five regions of phase space, evaluated using the MC simulation and smoothed.
The grey band displays the symmetrized envelope around the \ttbar FFs.
The envelope is used to estimate the systematic uncertainty of the \ttbar FFs.
The highest \pt bin contains events with fake-\tauhadvis with \pt of up to 300~\gev.
}
\label{fig:uff_ff_syst_b}
\end{figure}

\begin{figure}
\centering
\subfloat[]{
\includegraphics[width=0.45\textwidth]{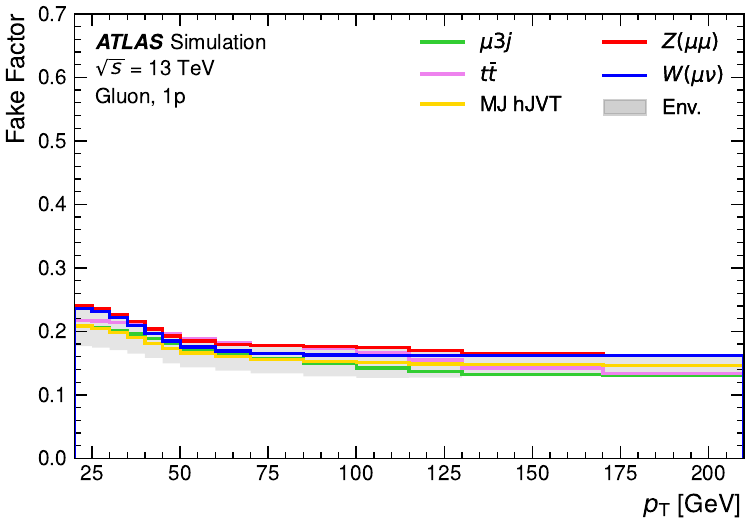}
\label{fig:uff_ff_syst_g_1p}
}\quad
\subfloat[]{
\includegraphics[width=0.45\textwidth]{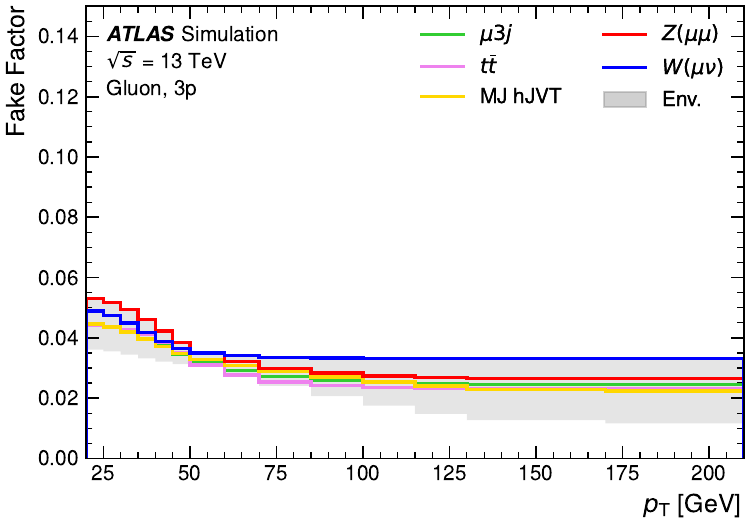}
\label{fig:uff_ff_syst_g_3p}
}
\caption{FFs for the (a) 1-prong (1p) \gfake fakes and (b) 3-prong (3p) \gfake fakes as a function of the fake-\tauhadvis \pt  in five regions of phase space, evaluated using the MC simulation and smoothed.
The grey band displays the symmetrized envelope around the \mjh FFs.
The envelope is used to estimate the systematic uncertainty of the \mjh FFs.
The highest \pt bin contains events with fake-\tauhadvis with \pt of up to 300~\gev.
}
\label{fig:uff_ff_syst_g}
\end{figure}

The systematic uncertainty of the \mjl FFs is estimated in a similar way.
The main difference is that data is used instead of MC simulation.
The JVT score of the jets seeding the \tauhadvis candidate is used to select suitable \pfake fake
samples from data.
Five regions are defined, including the \mjl and \gl regions, and the systematic uncertainty is determined as
the largest difference between the \mjl FF and the other FFs within each \pt and prongness bin.
The uncertainty is symmetrized.
Three regions, \Zmml, \ttbarl, and \wmunul are defined as the \Zmm, \ttbar, and \wmunu regions in which just the \jvt requirement
on the \tauhadvis candidate's seed jet is inverted: $\jvt < 0.8$ is required.
Figure~\ref{fig:uff_ff_syst_p} shows the FFs for the \pfake fakes in the five regions and the uncertainty of the \mjl FFs.
\begin{figure}
\centering
\subfloat[]{
\includegraphics[width=0.45\textwidth]{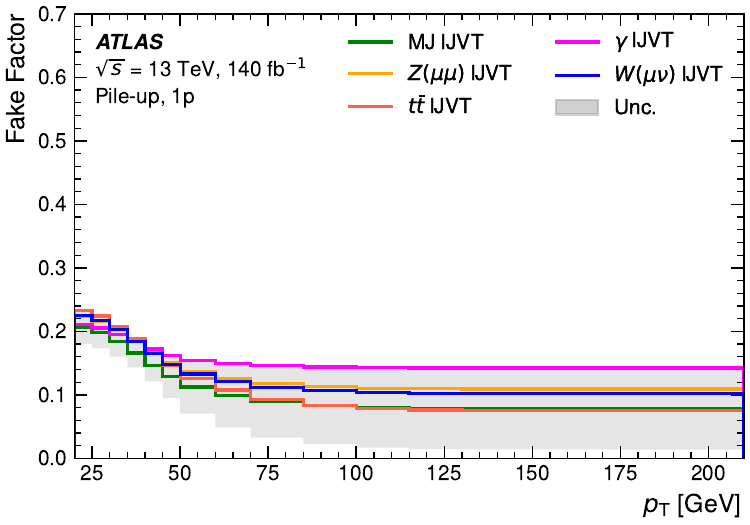}
\label{fig:uff_ff_syst_p_1p}
}\quad
\subfloat[]{
\includegraphics[width=0.45\textwidth]{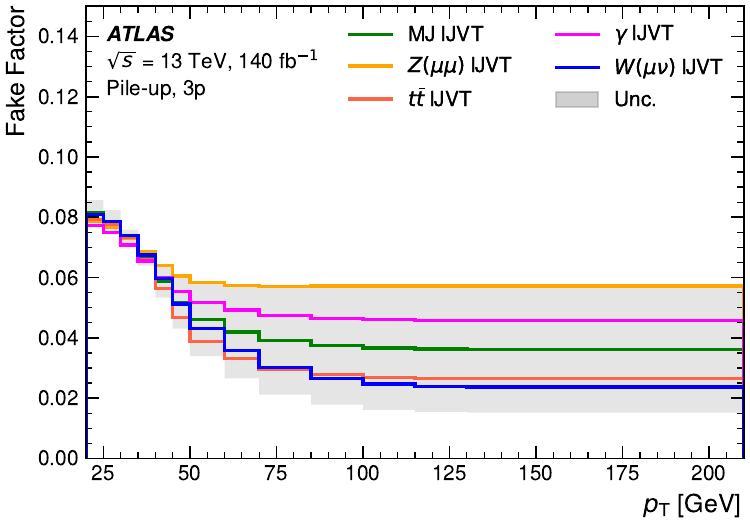}
\label{fig:uff_ff_syst_p_3p}
}
\caption{FFs for the (a) 1-prong (1p) \pfake fakes and (b) 3-prong (3p) \pfake fakes as a function of the fake-\tauhadvis \pt  in five regions of phase space and smoothed.
The grey band displays the assigned systematic uncertainty of the \mjl FFs.
The highest \pt bin contains events with fake-\tauhadvis with \pt of up to 300~\gev.
}
\label{fig:uff_ff_syst_p}
\end{figure}

The systematic uncertainty is propagated to \ffuff separately for each uncertainty source (i.e.\  the \qfake, \gfake, \bfake or \pfake FF uncertainty).
One FF from Eq.~(\ref{eq:uff}) is varied at a time, both upwards and downwards, and in each case the \ffuff value is recalculated, giving rise to the $\pm 1 \sigma$ \ffuff uncertainty estimate.
For each source, the resulting \ffuff uncertainty is smoothed as a function of \pt to reduce the impact of statistical fluctuations.
For the smoothing, the relative uncertainty is evaluated,
copied to a uniform binning with a bin width of 1
and smoothed using a Gaussian kernel with a width of 1.
Finally, it is copied back to the original \pt bins.
These uncertainty sources should be treated as being uncorrelated and introduced as four separate nuisance parameters for each prongness in statistical models, assuming a correlation between \pt bins and no correlation between prongness bins.
If an analysis is susceptible to the \pt and prongness correlation scenario,
it can determine whether a specific decorrelation in \pt and/or correlation in prongness results in a more realistic model.
Decorrelation in other variables can also be considered.
All uncertainty components are summed in quadrature for display purposes.

Additional systematic effects would arise if the q/g score spectra of \qgbp fakes were different in different regions of phase space.
The spectra were studied using MC (data) for \qgb (\pfake) fakes, exploiting the same regions as in the FFs comparison.
Some differences larger than the statistical uncertainty were observed among the \qfake fake templates;
see Figure~\ref{fig:uff_q_shapes} as an example.
The figure compares the distributions of 1p \qfake fakes with $\pt \in (20, 35)~\gev$
determined in the \Zmm, \ttbar, \mjh, \wmunu, and \muthreej antiID subregions
using the same MC samples as for the systematic uncertainty determination described above.
A MC study was performed to estimate the impact of this effect on the \ffuff value.
MC templates for the \gfake, \bfake, and \pfake fakes are built in the \Zmm antiID subregion,
using the \Zmm MC sample from the Powheg set.
Similarly, MC templates for the \qfake fakes are built in the \wmunu antiID subregion,
using the \wmunu MC sample from the Powheg set.
Each template is normalized to the total fake yield
in the corresponding \pt and prongness bin of the \Zmm antiID subregion.
FFs for the \qgbp fakes are determined in the \Zmm region
using the \Zmm MC sample from the Powheg set.
In each \pt and prongness bin, the \qgbp templates are fitted to the total fake yield in the \Zmm antiID subregion, similarly to the procedure described in Section~\ref{sec:uff_fit}.
The FFs are then combined in a similar way as described in Section~\ref{sec:uff_idea}.
The resulting FF is compared to a reference FF directly calculated using \Zmm fakes predicted by the \Zmm MC sample from the Powheg set.
In most \pt and prongness bins, the difference between the resulting FF and the \Zmm one is within their statistical uncertainty.
In the minority of the bins, the difference is larger than the statistical uncertainty but within the systematic uncertainty of the \qfake FFs, see Figure~\ref{fig:uff_ff_syst_q}.
Therefore, no additional systematic uncertainty is assigned to the \ffuff value due to the differences in the q/g score spectra of \qgbp fakes in different regions of phase space.
\begin{figure}
\centering
\includegraphics[width=0.45\textwidth]{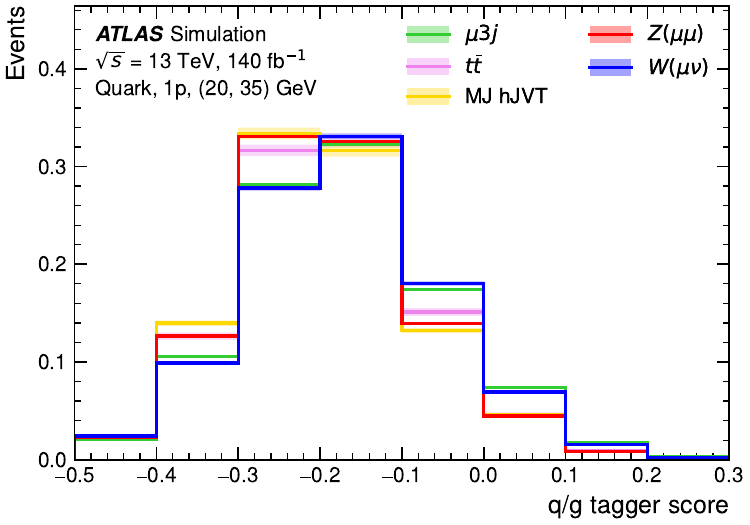}
\caption{
Spectra of the q/g tagger score determined for \qfake fakes in the \Zmm, \ttbar, \mjh, \wmunu, and \muthreej antiID subregions using MC.
They are normalized to unity.
The bands correspond to the statistical uncertainty of each spectrum.
}
\label{fig:uff_q_shapes}
\end{figure}

It was also found that varying the JVT cut value from $0.8$ to $0.5$ or $0.9$ in the
\Zmm, \ttbar, \mjh and \mjl region definitions does not affect the results significantly.


%

%

\section{Tests of the Universal Fake Factor method}
\label{sec:validation}

The UFF method is tested in several regions, and
the test in the \wmunu region is discussed here in depth as an example.
Given that the region is highly enriched in fakes, it is possible to determine the FFs directly -- as the ratio of the ID to antiID event yields.
This gives a reference value for the \ffuff that are evaluated using the q/g tagger score in the antiID subregion in order to predict the ID subregion yields.
The FFs are compared in Figure~\ref{fig:ff_wmunu}.
The systematic uncertainty of the estimated \ffuff ranges between 15\% and 35\%, depending on \pt and prongness.
The UFF predictions agree with the reference FFs within the uncertainties.
However, the \ffuff are shifted systematically upwards relative the reference FFs.
This is due to the \qfake FFs being lower in the \wmunu region
than in the \Zmm and \mjh regions that are used
in the combination formula of Eq.~(\ref{eq:uff}).
Figure~\ref{fig:uff_ff_syst_q} shows the various FFs estimated with the MC simulation.
\begin{figure}
\centering
\subfloat[]{
\includegraphics[width=0.45\textwidth]{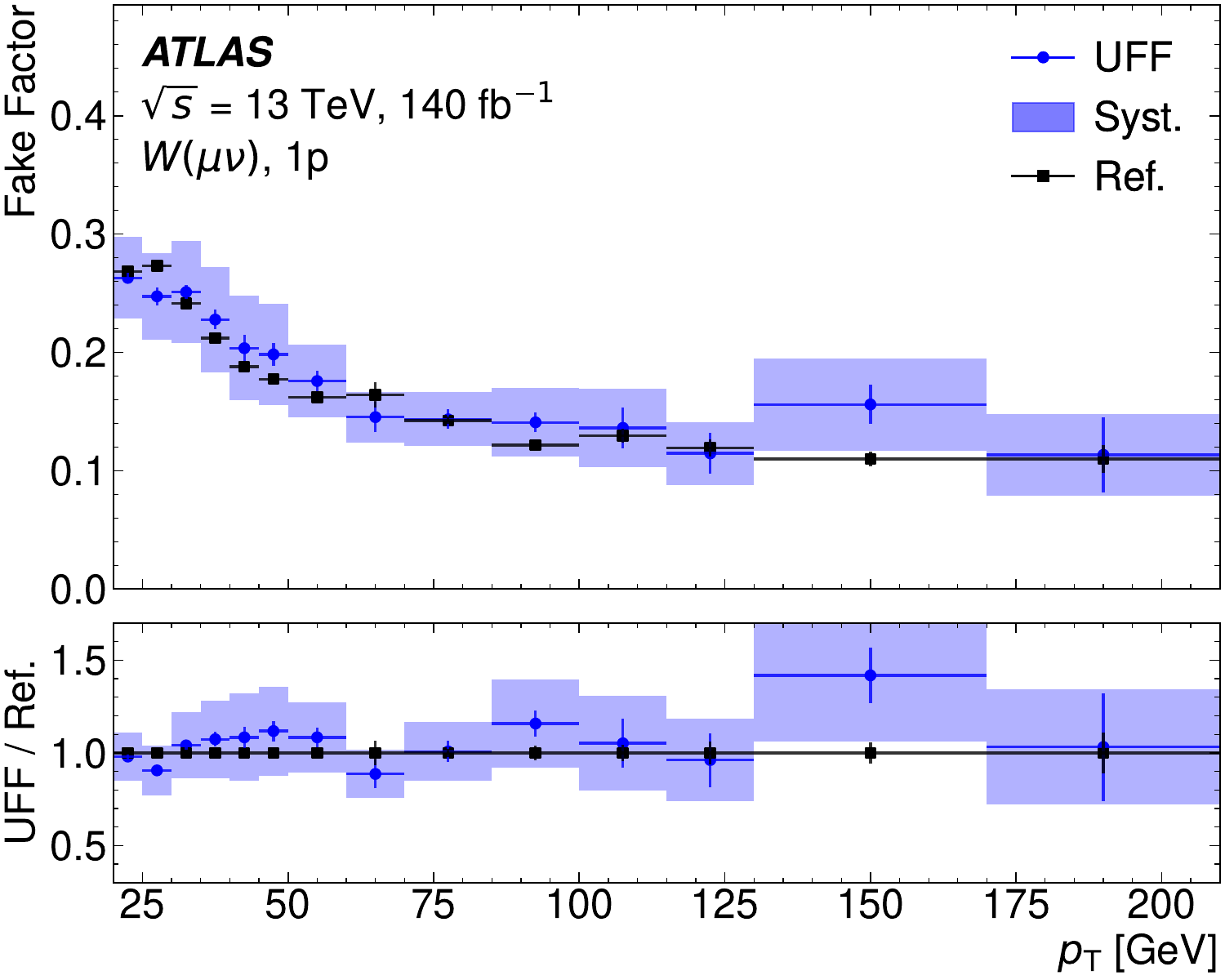}
\label{fig:ff_wmunu_1p}
}\quad
\subfloat[]{
\includegraphics[width=0.45\textwidth]{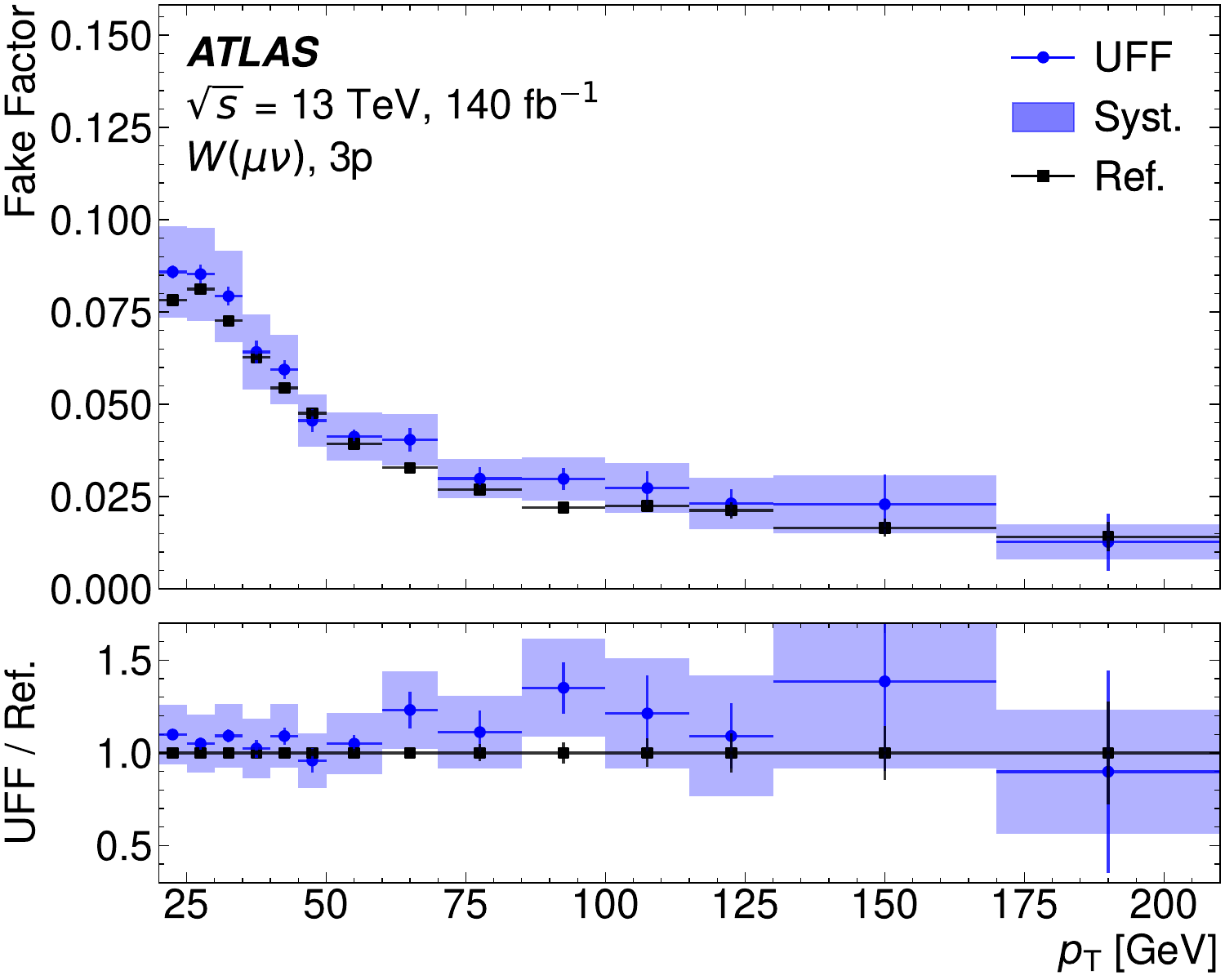}
\label{fig:ff_wmunu_3p}
}
\caption{FFs for the (a) 1-prong (1p) and (b) 3-prong (3p) fakes as a function of the fake-\tauhadvis \pt in the \wmunu region.
Blue-circle points display FFs estimated using the UFF method.
The statistical (systematic) uncertainty is displayed as the blue error bars (band).
The reference FFs, which are measured directly in the region, are shown as the black-square points,
and their statistical uncertainty is displayed as the black error bars.
The highest \pt bin contains events with fake-\tauhadvis with \pt of up to 300~\gev.
}
\label{fig:ff_wmunu}
\end{figure}

To test the UFF method further, the fake-background prediction is compared with
the data in the \wmunu ID subregion.
Figure~\ref{fig:spectra_wmunu_1p} (Figure~\ref{fig:spectra_wmunu_3p}) shows the comparison for events containing a
1-prong (3-prong) \tauhadvis candidate, and for several variables, namely the \tauhadvis transverse momentum and $\eta$,
the muon transverse momentum \ptmu, the \met, the transverse mass \mumetmt of the muon and \met system, and
the angular separation \mutaudr of the \tauhadvis and muon.
The real-\tl-lepton background is also shown.
It is estimated using the Default set of MC samples.
The fake-background prediction agrees with the data, within the uncertainties, in all the variables.
The systematic shift of the \ffuff discussed above is also visible in this comparison.
\begin{figure}
\centering
\subfloat[]{
\includegraphics[width=0.32\textwidth]{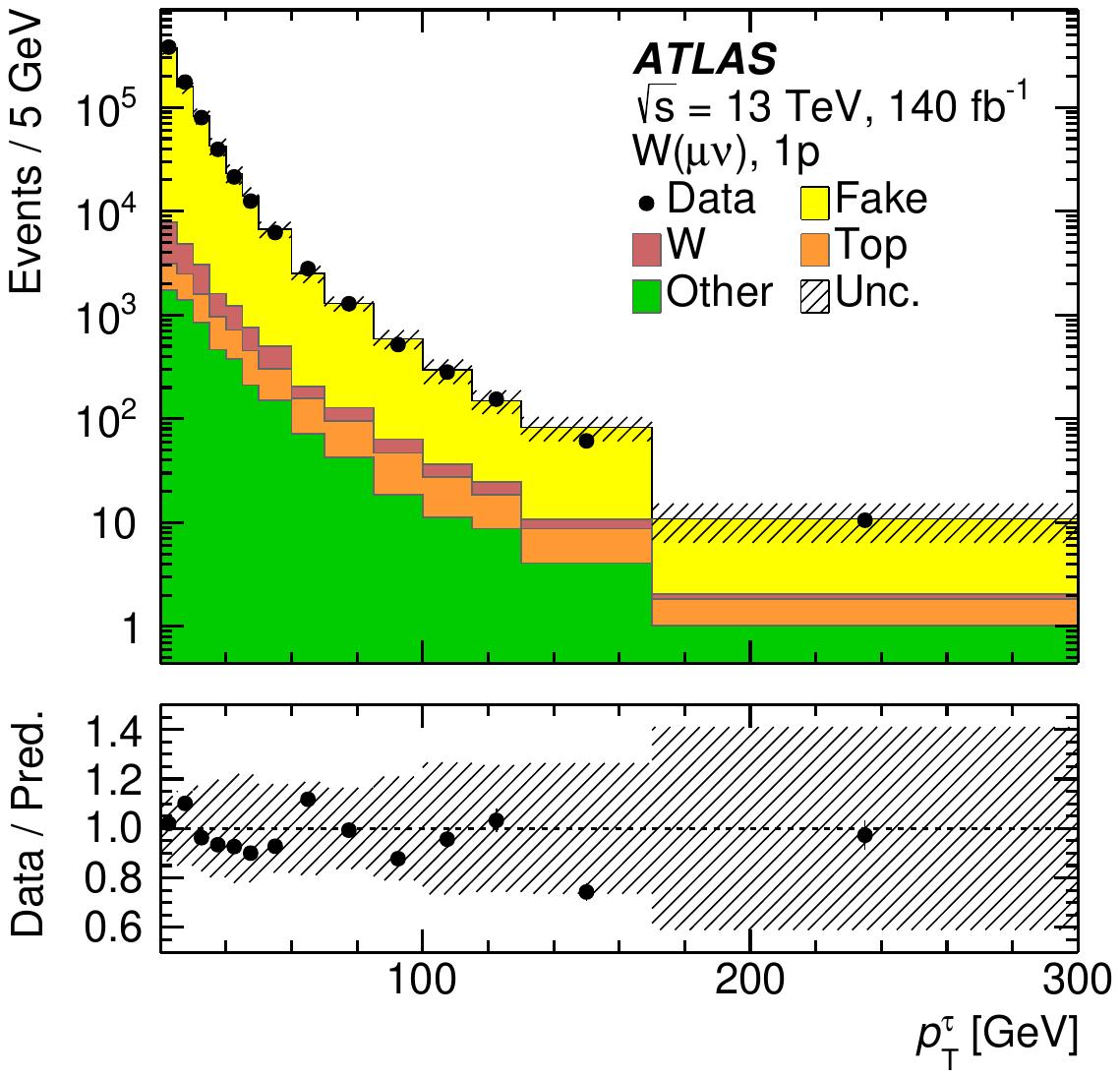}
\label{fig:spectra_wmunu_1p_tau0_pt}
}\quad
\subfloat[]{
\includegraphics[width=0.32\textwidth]{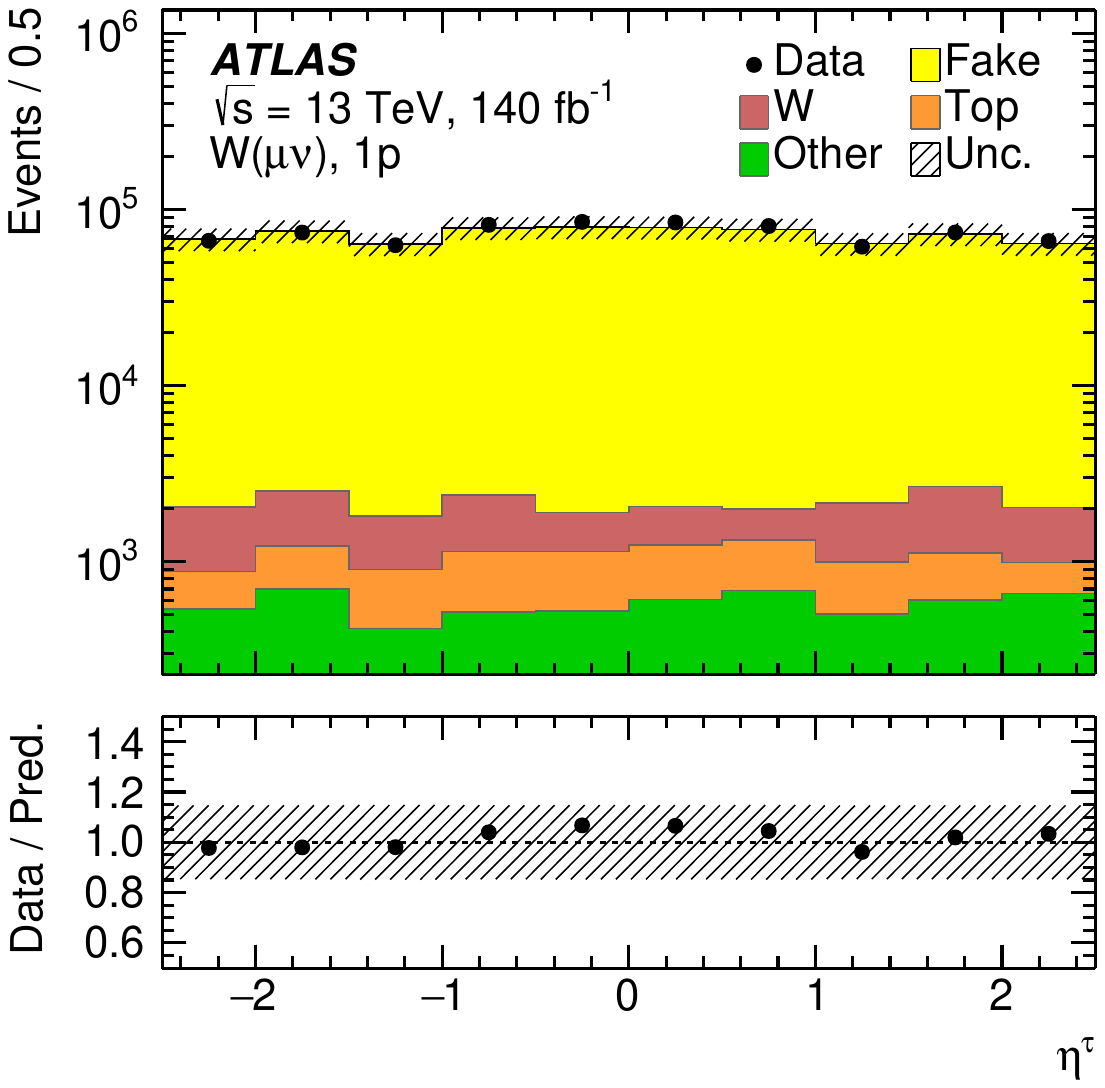}
\label{fig:spectra_wmunu_1p_tau0_eta}
}\quad
\subfloat[]{
\includegraphics[width=0.32\textwidth]{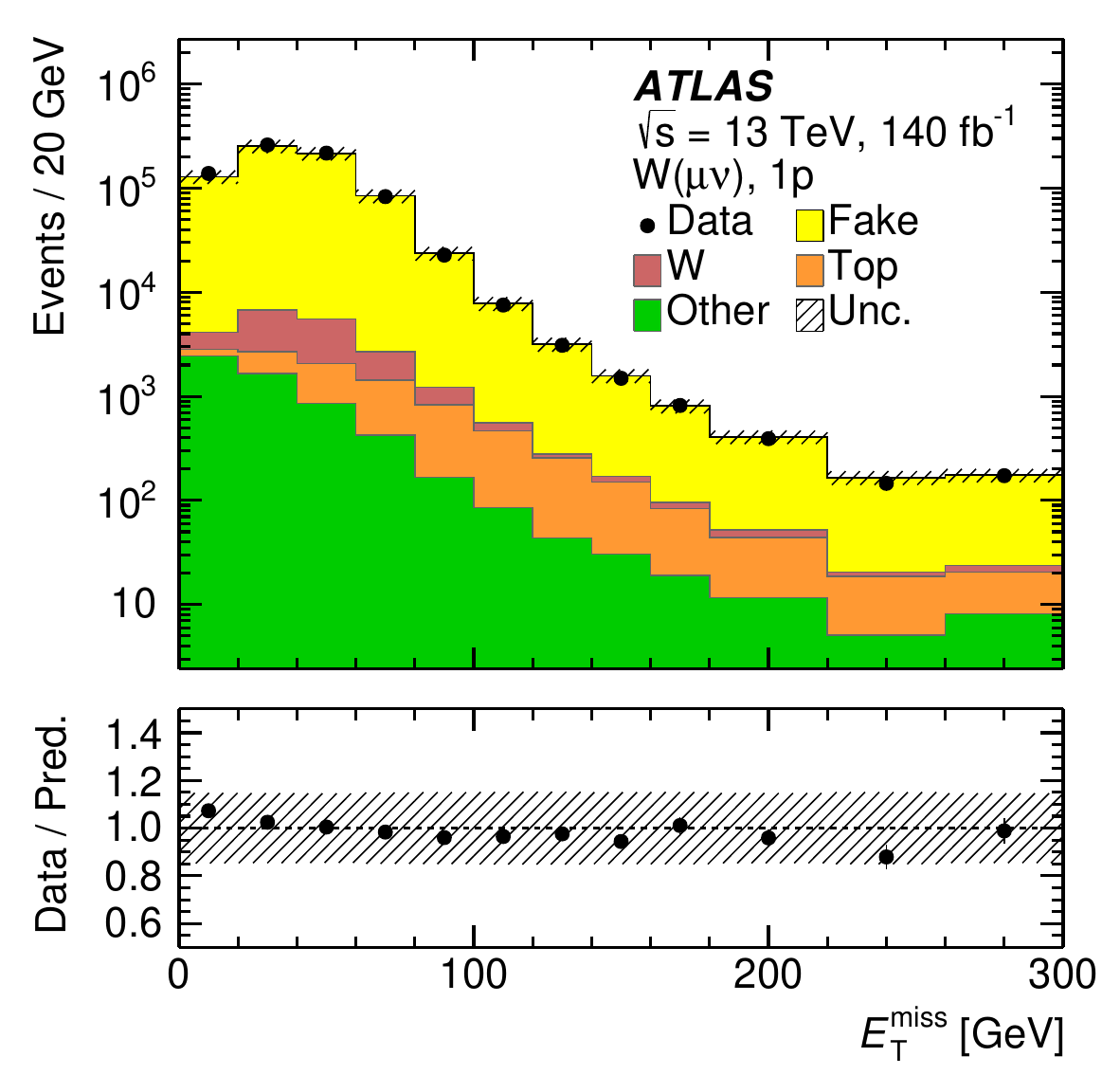}
\label{fig:spectra_wmunu_1p_met}
}\quad
\subfloat[]{
\includegraphics[width=0.32\textwidth]{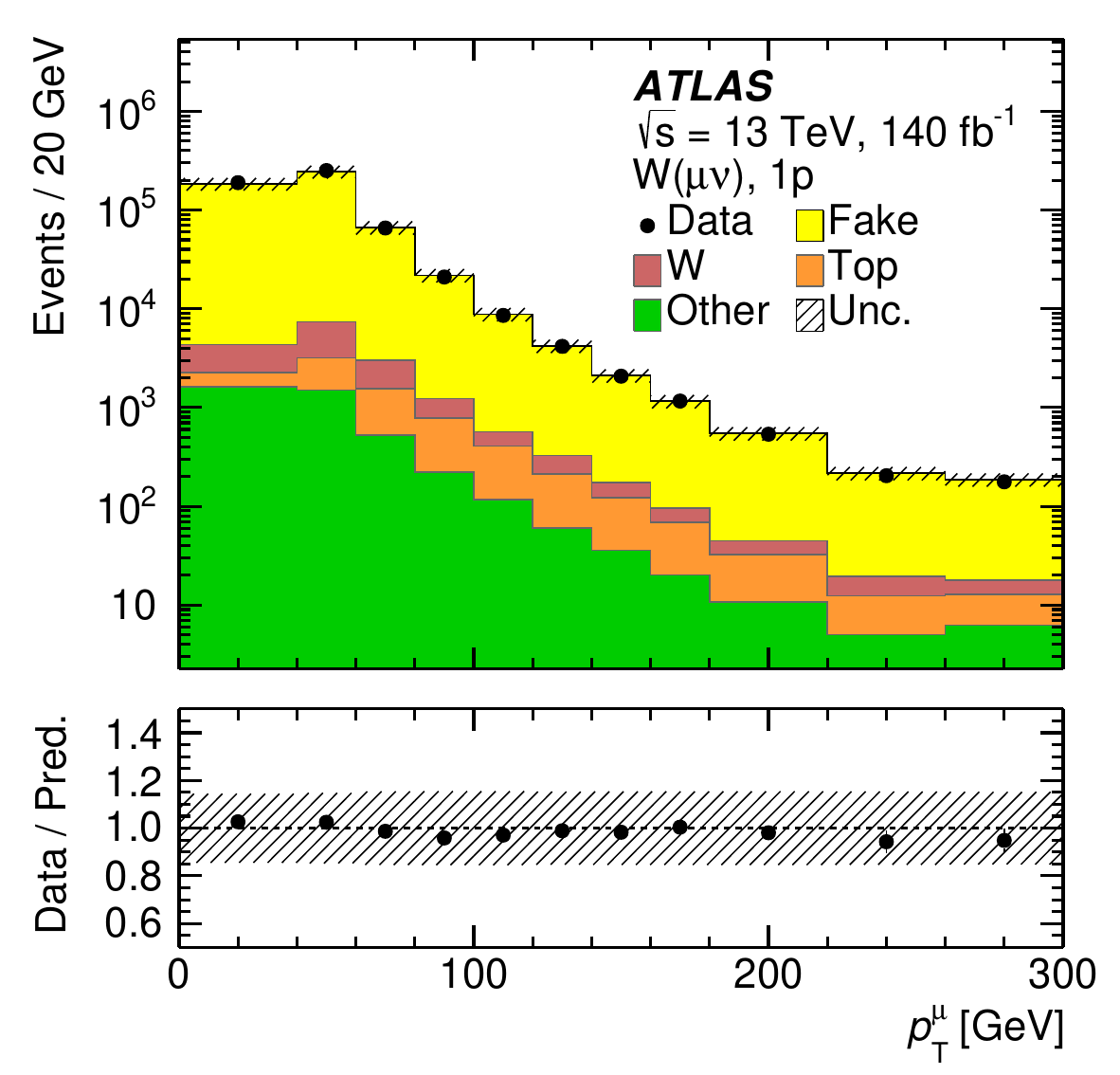}
\label{fig:spectra_wmunu_1p_mu0_pt}
}\quad
\subfloat[]{
\includegraphics[width=0.32\textwidth]{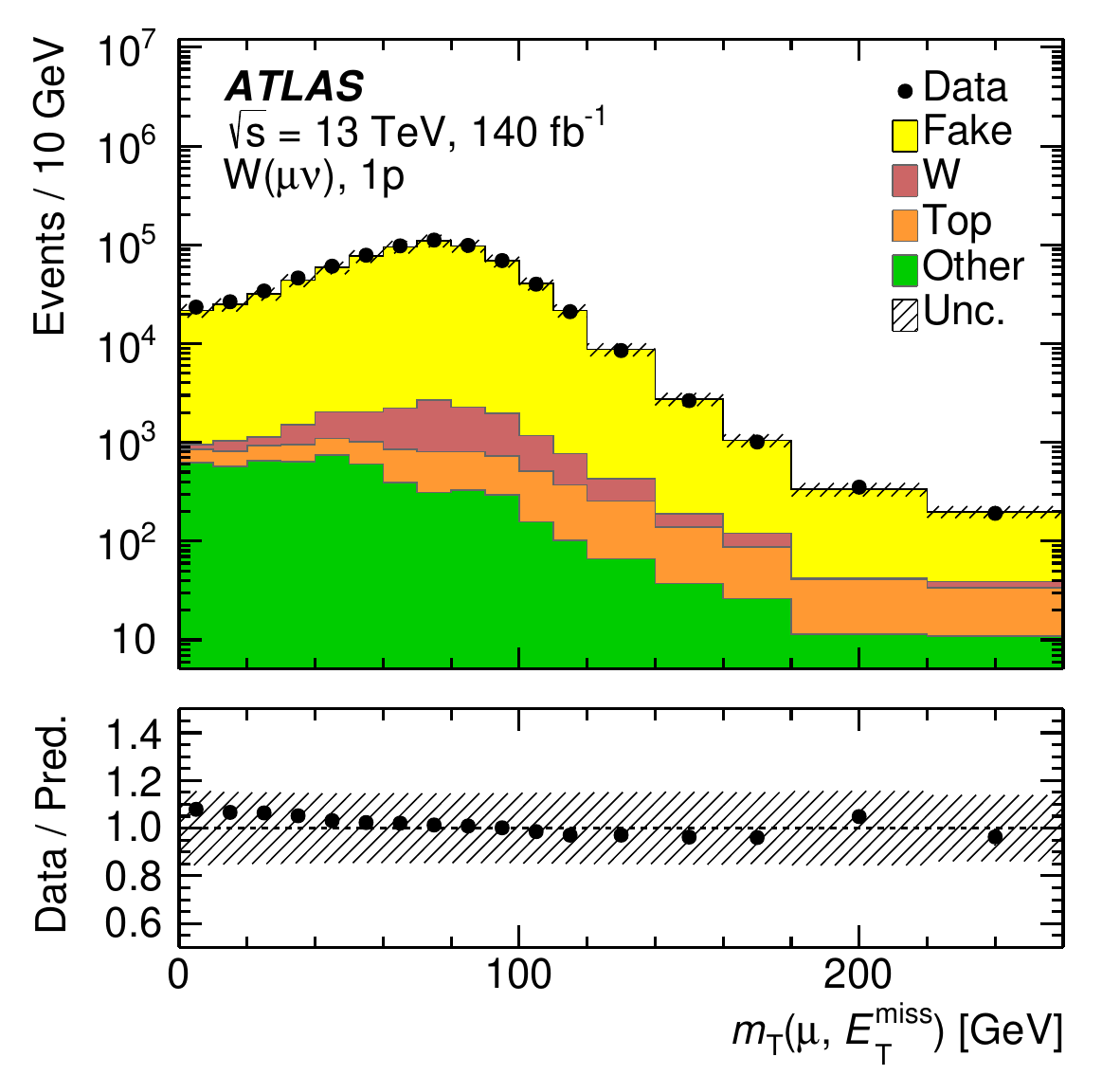}
\label{fig:spectra_wmunu_1p_mu_met_mt}
}\quad
\subfloat[]{
\includegraphics[width=0.32\textwidth]{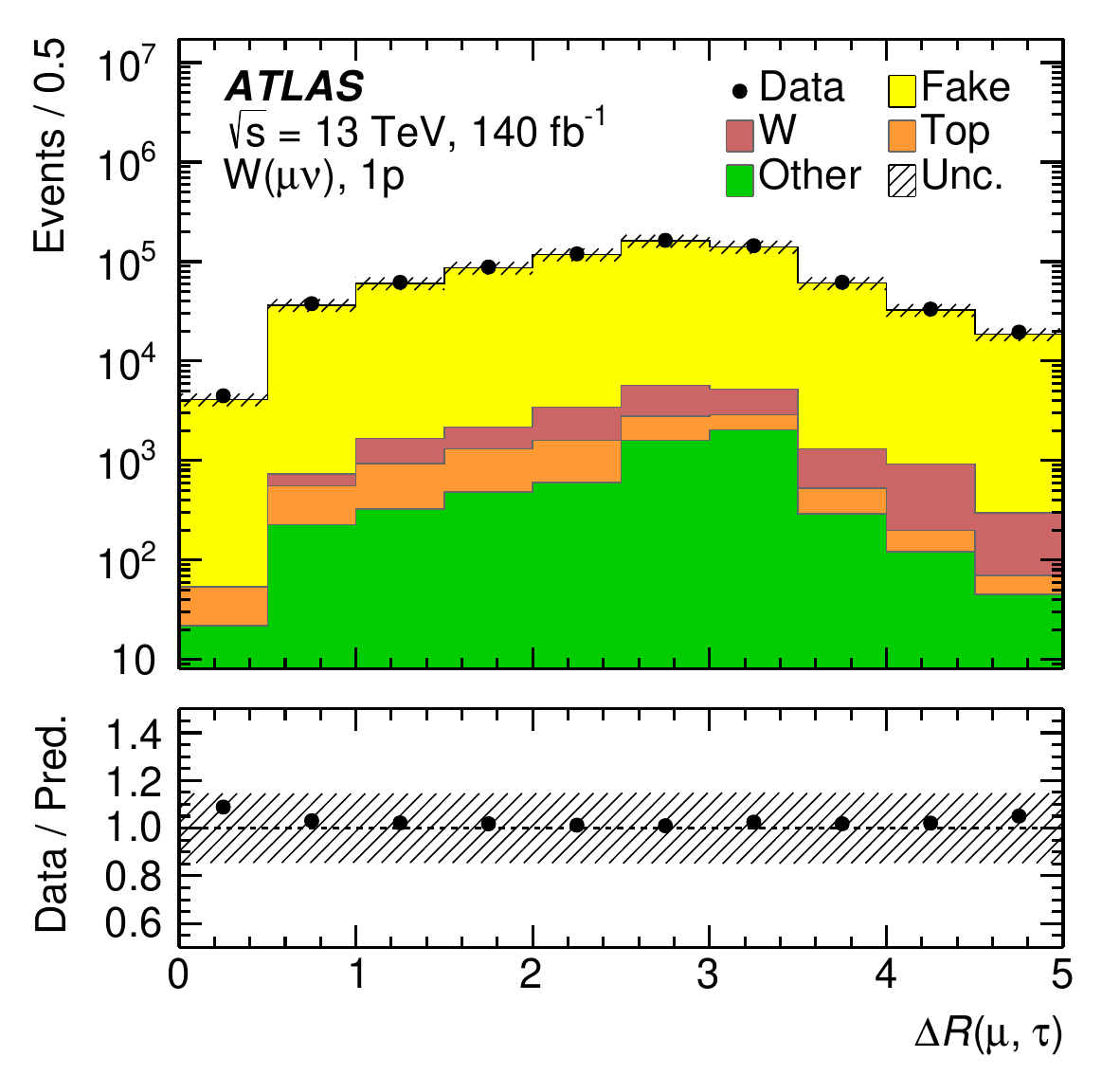}
\label{fig:spectra_wmunu_1p_mu_tau_dR}
}
\caption{
Comparison of the 1-prong (1p) fake-background prediction (yellow area) with the data (black points) in the \wmunu region.
The real-\tl-lepton backgrounds are also shown.
The shaded band displays the total uncertainty of the prediction.
It includes the fake-related systematic uncertainty, the statistical uncertainty of the \ffuff,
and the uncertainty due to the limited number of events in the antiID subregion of the SR and the MC samples.
The comparison is shown for several variables: (a) \tauhadvis transverse momentum, (b) \tauhadvis $\eta$, (c) \met, (d) \ptmu, (e) \mumetmt, and (f) \mutaudr.
The compact notation \pttau and \etatau is used in the $x$-axis labels of figures (a) and (b) to
denote the transverse momentum and $\eta$ of \tauhadvis.
}
\label{fig:spectra_wmunu_1p}
\end{figure}

\begin{figure}
\centering
\subfloat[]{
\includegraphics[width=0.32\textwidth]{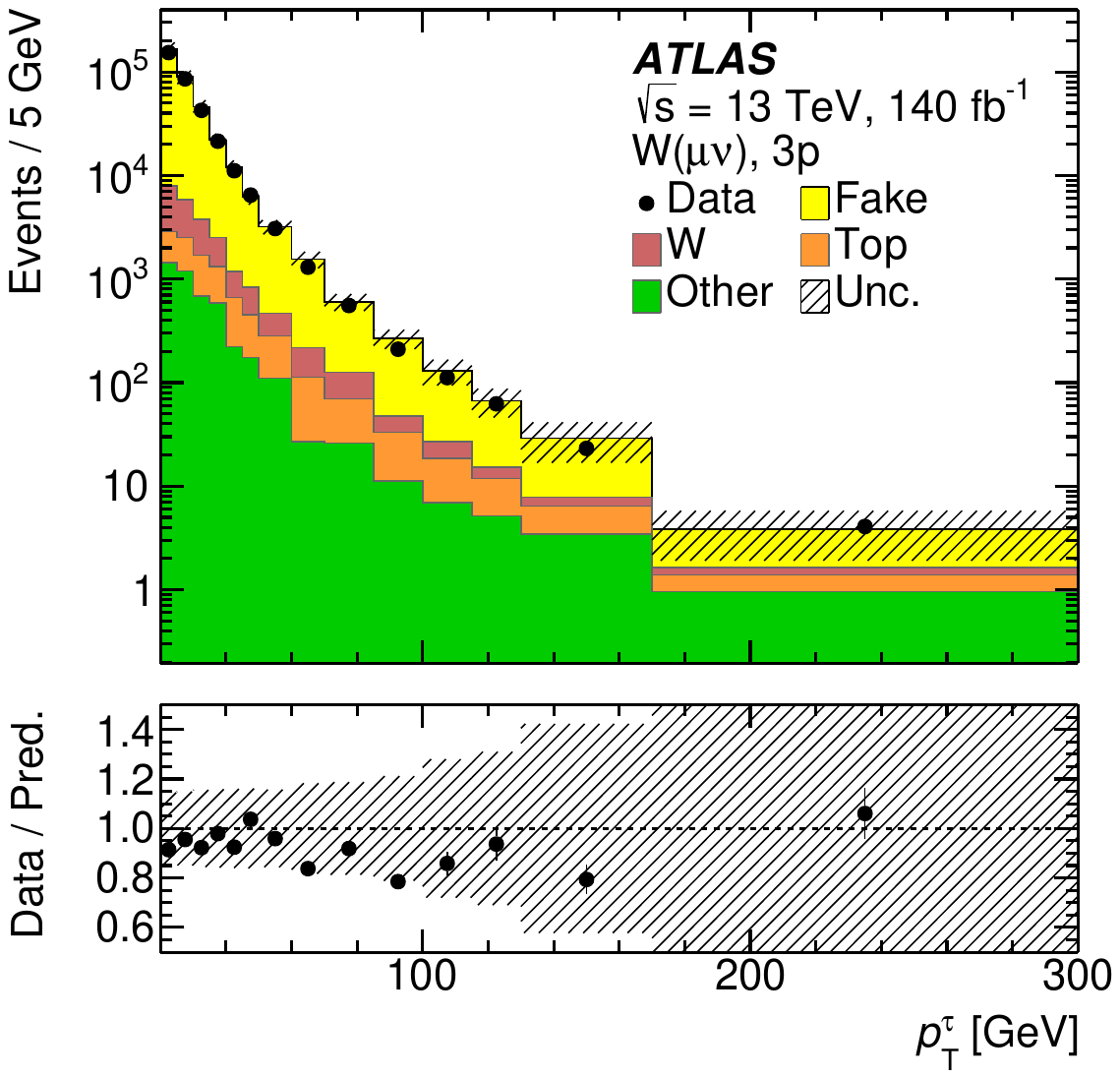}
\label{fig:spectra_wmunu_3p_tau0_pt}
}\quad
\subfloat[]{
\includegraphics[width=0.32\textwidth]{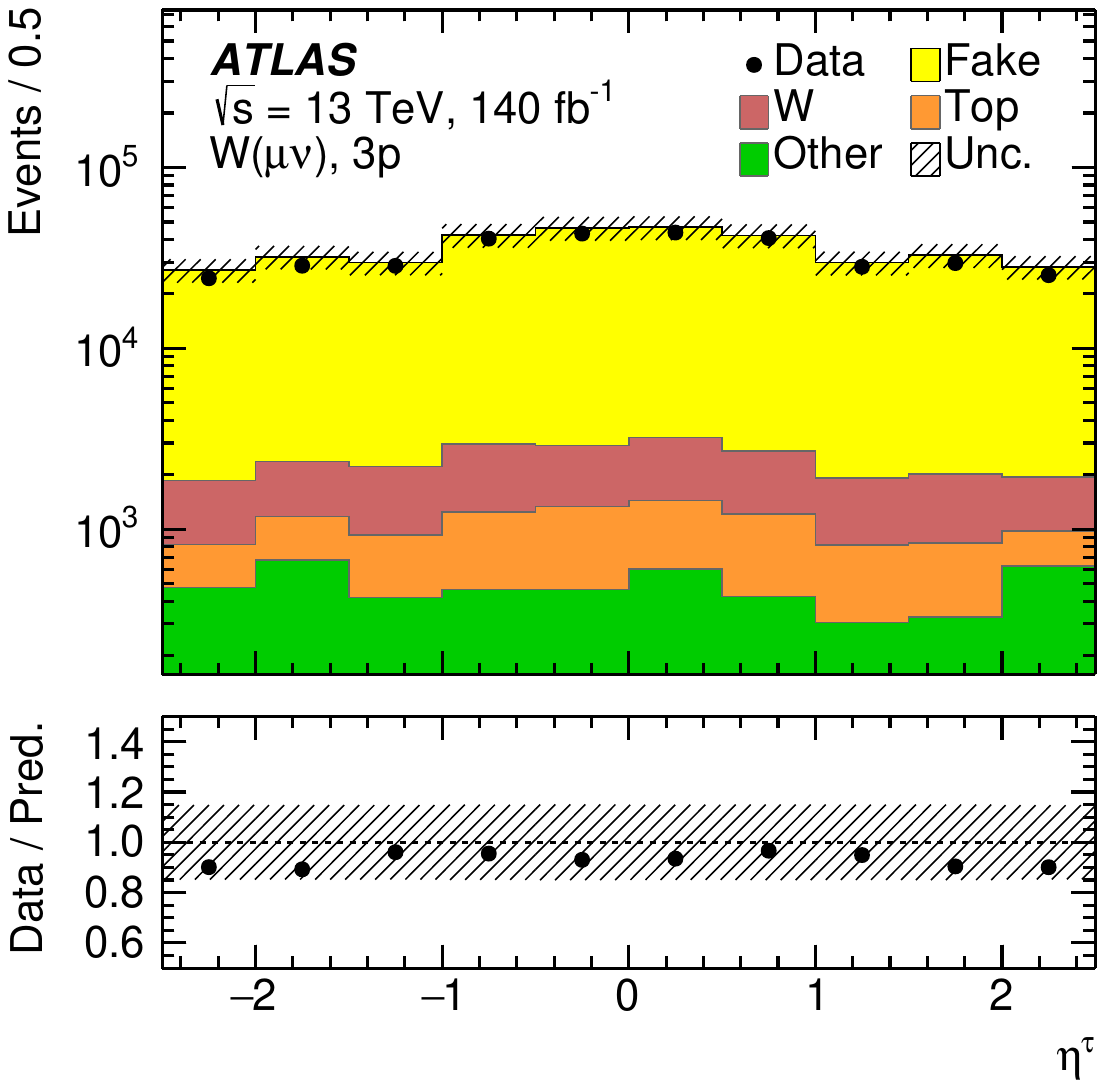}
\label{fig:spectra_wmunu_3p_tau0_eta}
}\quad
\subfloat[]{
\includegraphics[width=0.32\textwidth]{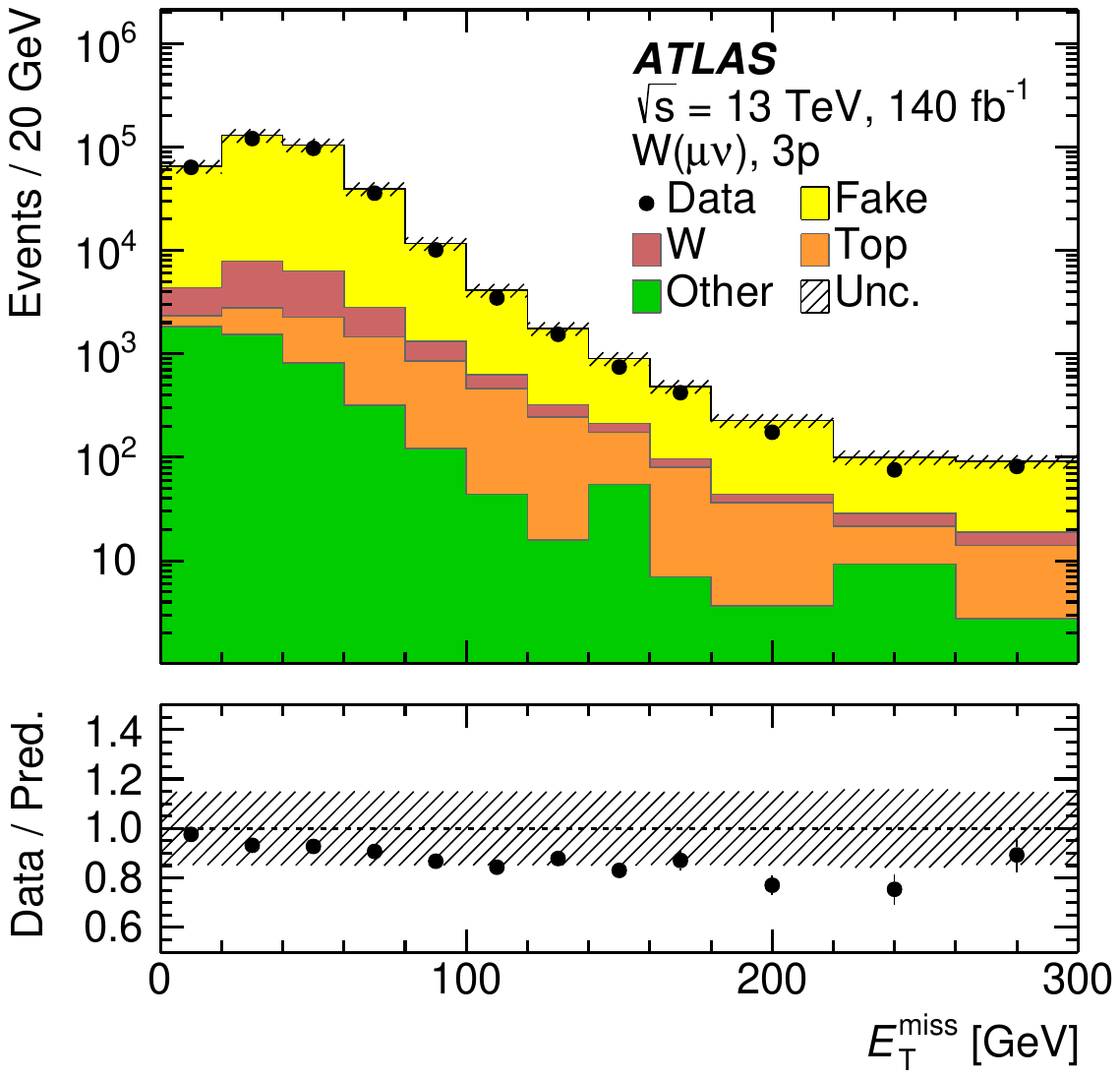}
\label{fig:spectra_wmunu_3p_met}
}\quad
\subfloat[]{
\includegraphics[width=0.32\textwidth]{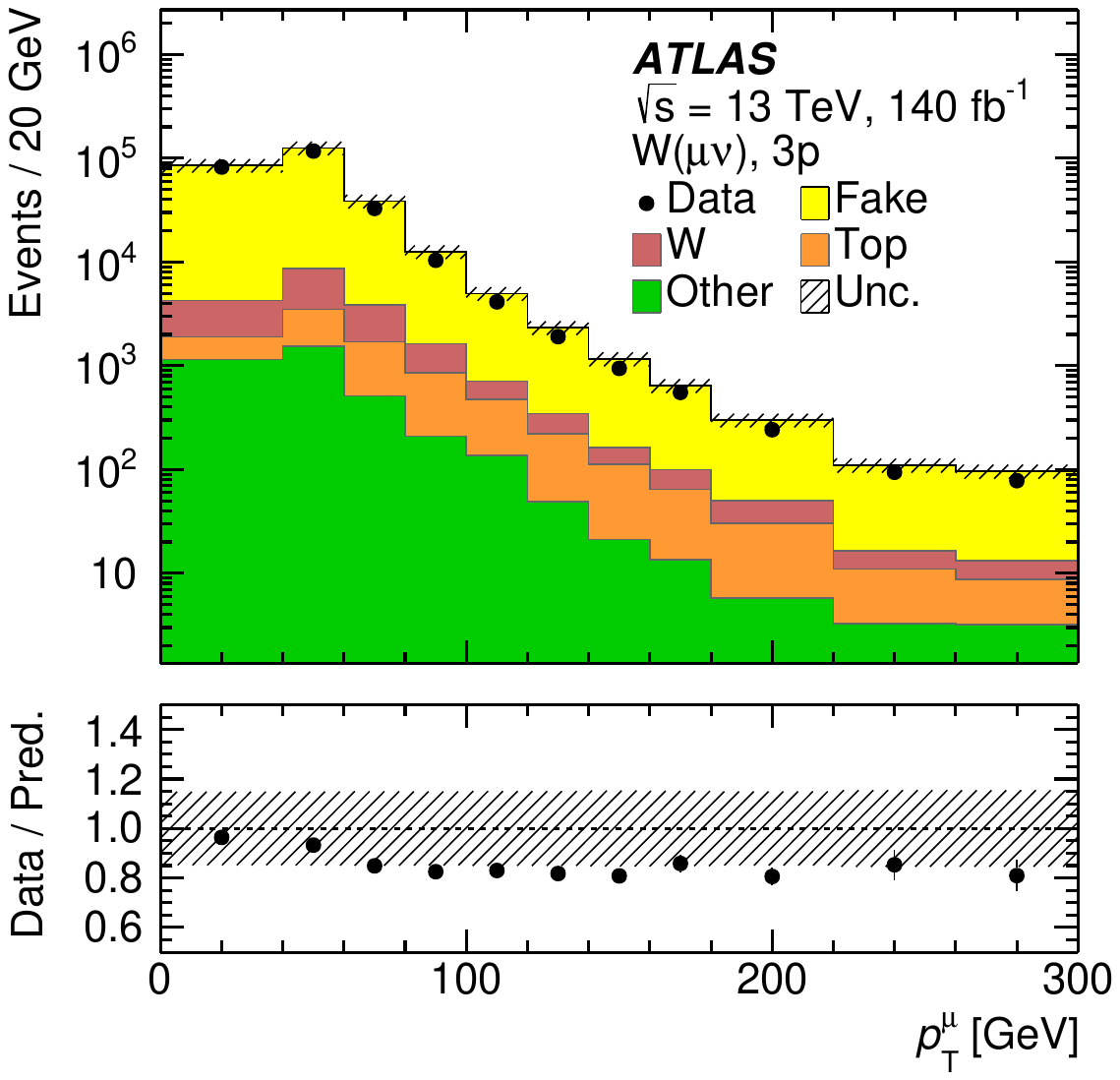}
\label{fig:spectra_wmunu_3p_mu0_pt}
}\quad
\subfloat[]{
\includegraphics[width=0.32\textwidth]{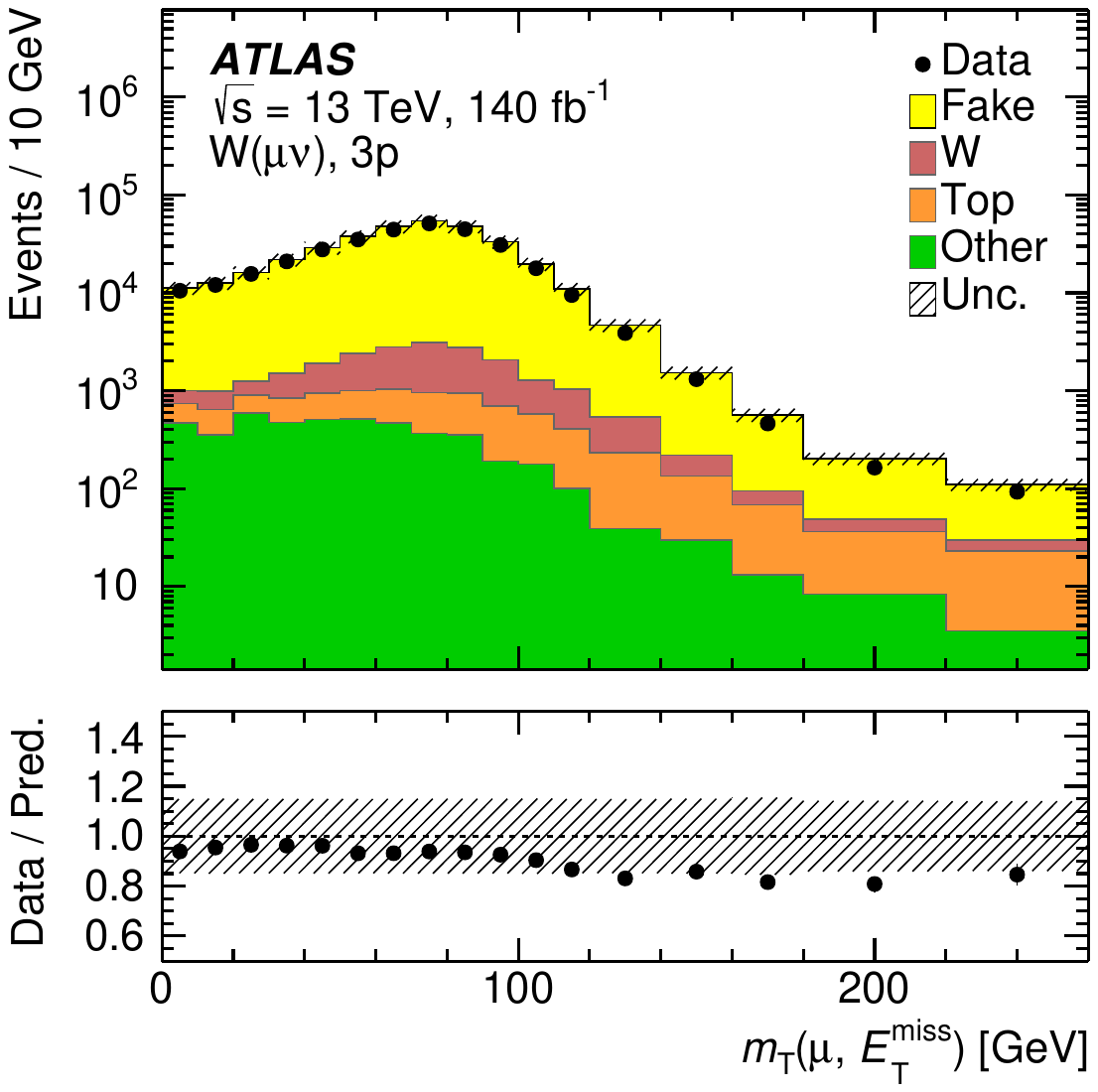}
\label{fig:spectra_wmunu_3p_mu_met_mt}
}\quad
\subfloat[]{
\includegraphics[width=0.32\textwidth]{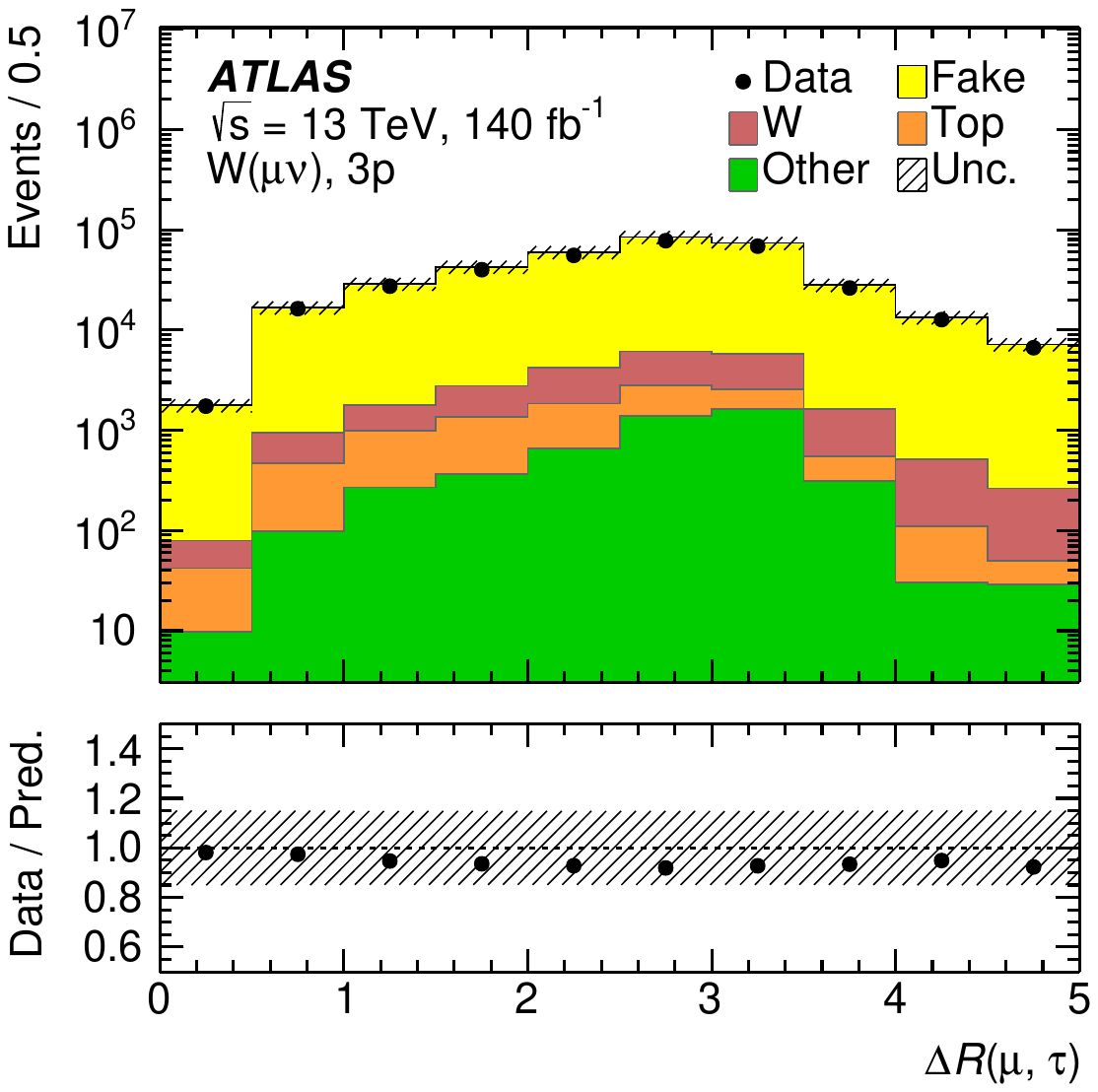}
\label{fig:spectra_wmunu_3p_mu_tau_dR}
}
\caption{
Comparison of the 3-prong (3p) fake-background prediction (yellow area) with the data (black points) in the \wmunu region.
The real-\tl-lepton backgrounds are also shown.
The shaded band displays the total uncertainty of the prediction.
It includes the fake-related systematic uncertainty, the statistical uncertainty of the \ffuff,
and the uncertainty due to the limited number of events in the antiID subregion of the SR and the MC samples.
The comparison is shown for several variables: (a) \tauhadvis transverse momentum, (b) \tauhadvis $\eta$, (c) \met, (d) \ptmu, (e) \mumetmt, and (f) \mutaudr.
The compact notation \pttau and \etatau is used in the $x$-axis labels of figures (a) and (b) to
denote the transverse momentum and $\eta$ of \tauhadvis.
}
\label{fig:spectra_wmunu_3p}
\end{figure}

The UFF method was also tested in \muthreej region,
where the fake composition is different from that in the \wmunu region, as shown in Figure~\ref{fig:uff_composition}.
The testing procedure in the \muthreej region is similar to that in \wmunu.
The \ffuff are compared with the reference FFs in Figure~\ref{fig:ff_mu3j}.
Figures~\ref{fig:spectra_mu3j_1p} and~\ref{fig:spectra_mu3j_3p} compare the fake-background
predictions with the data.
The UFF predictions agree with the data within the uncertainties.
Here, the \ffuff are shifted systematically downwards relative to the reference FFs, and
this is also visible in the comparison of the fake-background prediction with the data.
The reason is the same as discussed for the \wmunu region.
The different direction of the shift is a consequence of negative \Zmm normalization factors in the \muthreej region
and the fact that the \Zmm FFs are the highest ones, as shown in Figure~\ref{fig:uff_ff}.
\begin{figure}
\centering
\subfloat[]{
\includegraphics[width=0.45\textwidth]{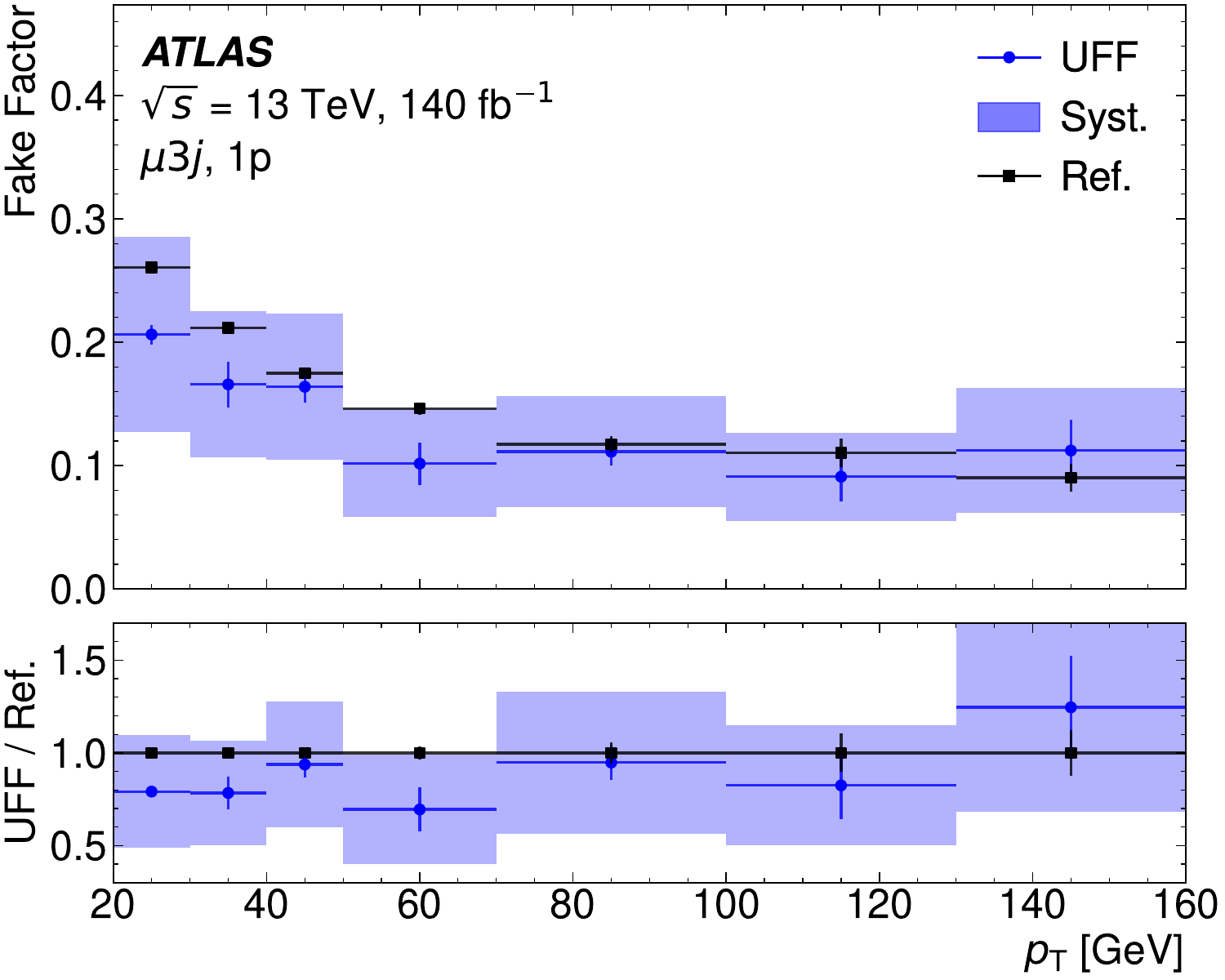}
\label{fig:ff_mu3j_1p}
}\quad
\subfloat[]{
\includegraphics[width=0.45\textwidth]{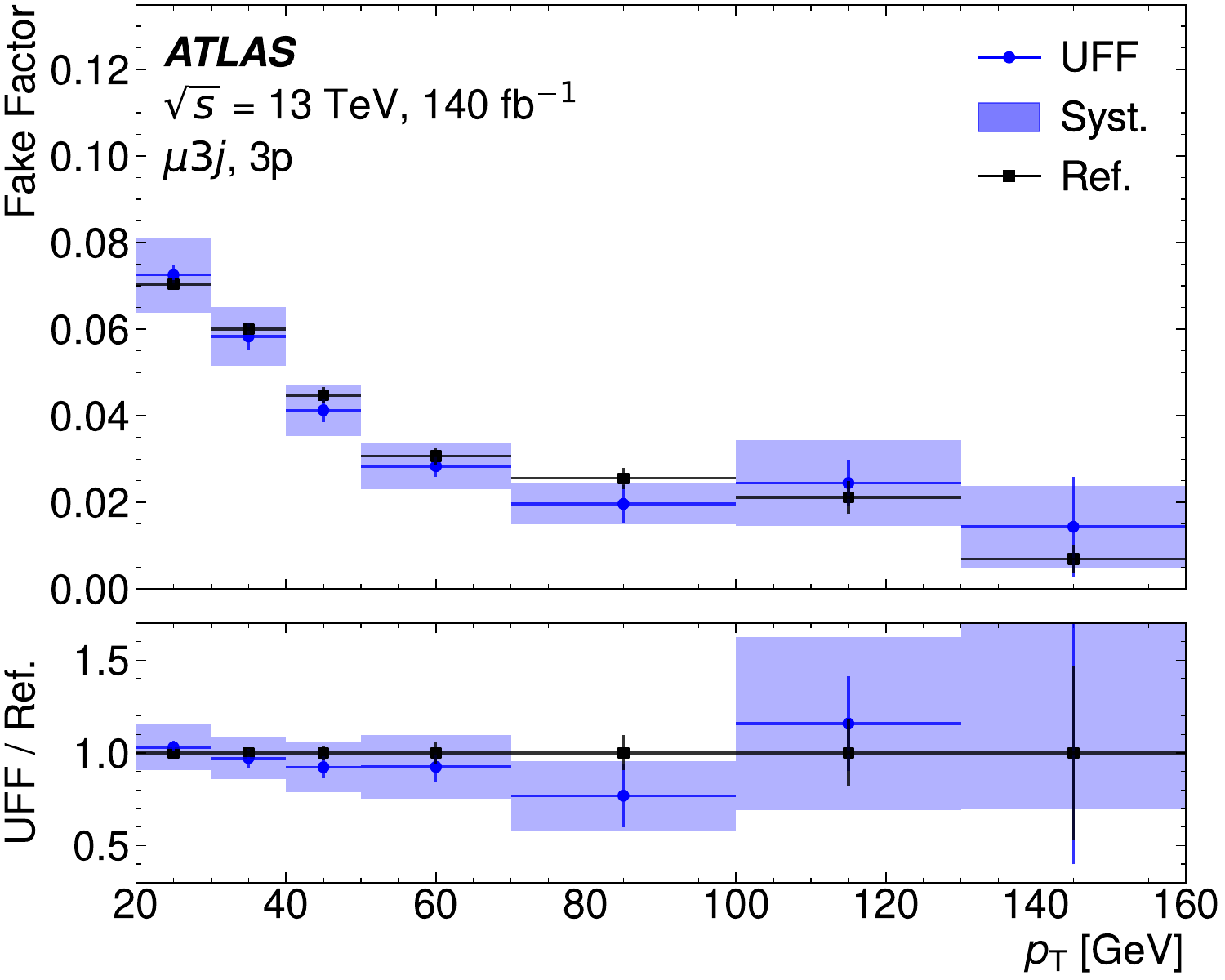}
\label{fig:ff_mu3j_3p}
}
\caption{FFs for the (a) 1-prong (1p) and (b) 3-prong (3p) fakes as a function of the fake-\tauhadvis \pt in the \muthreej region.
Blue-circle points display FFs estimated using the UFF method.
The statistical (systematic) uncertainty is displayed as the blue error bars (band).
The reference FFs, which are measured directly in the region, are shown as the black-square points,
and their statistical uncertainty is displayed as the black error bars.
The highest \pt bin contains events with fake-\tauhadvis with \pt of up to 300~\gev.
}
\label{fig:ff_mu3j}
\end{figure}
\begin{figure}
\centering
\subfloat[]{
\includegraphics[width=0.32\textwidth]{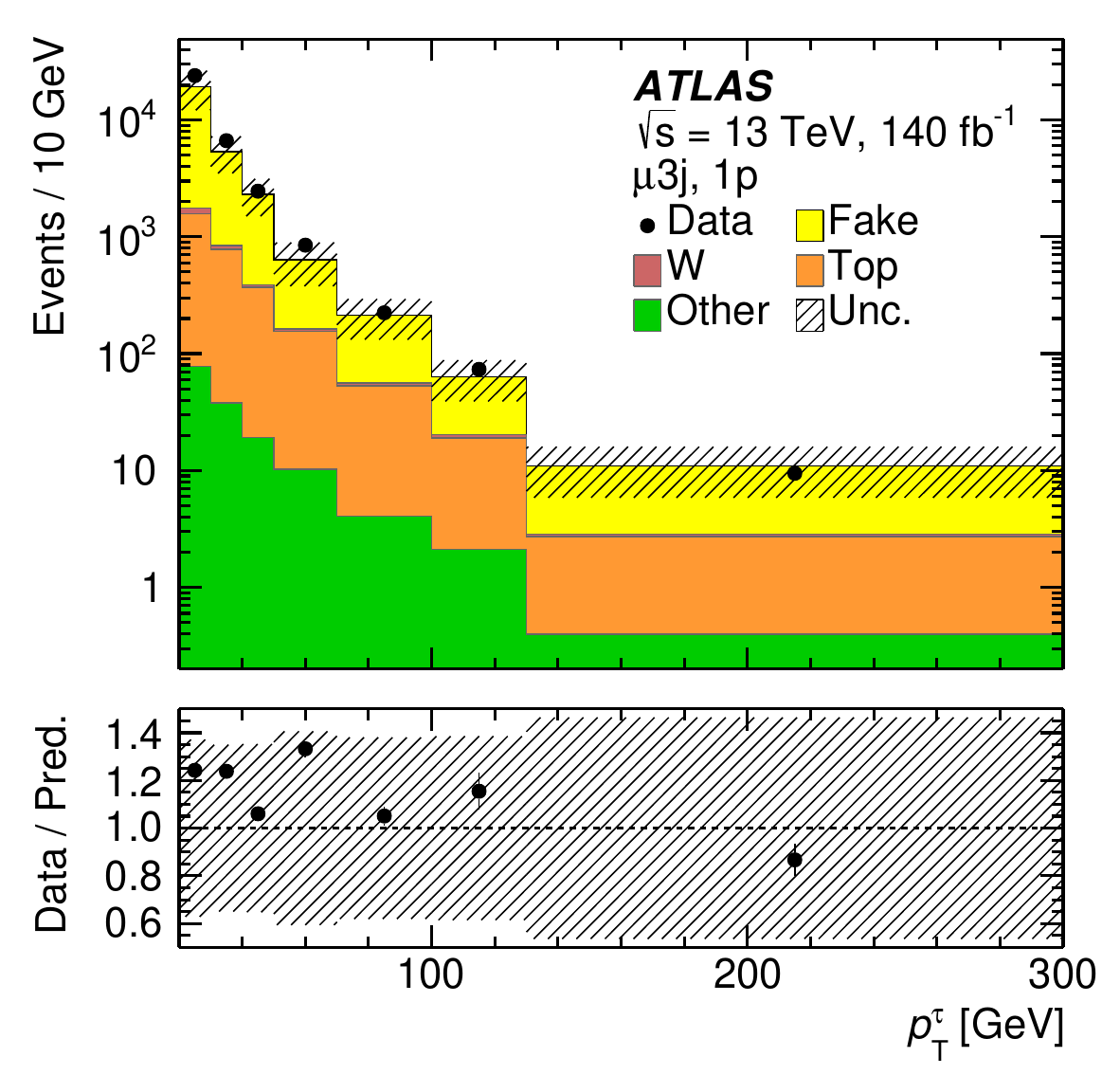}
\label{fig:spectra_mu3j_1p_tau0_pt}
}\quad
\subfloat[]{
\includegraphics[width=0.32\textwidth]{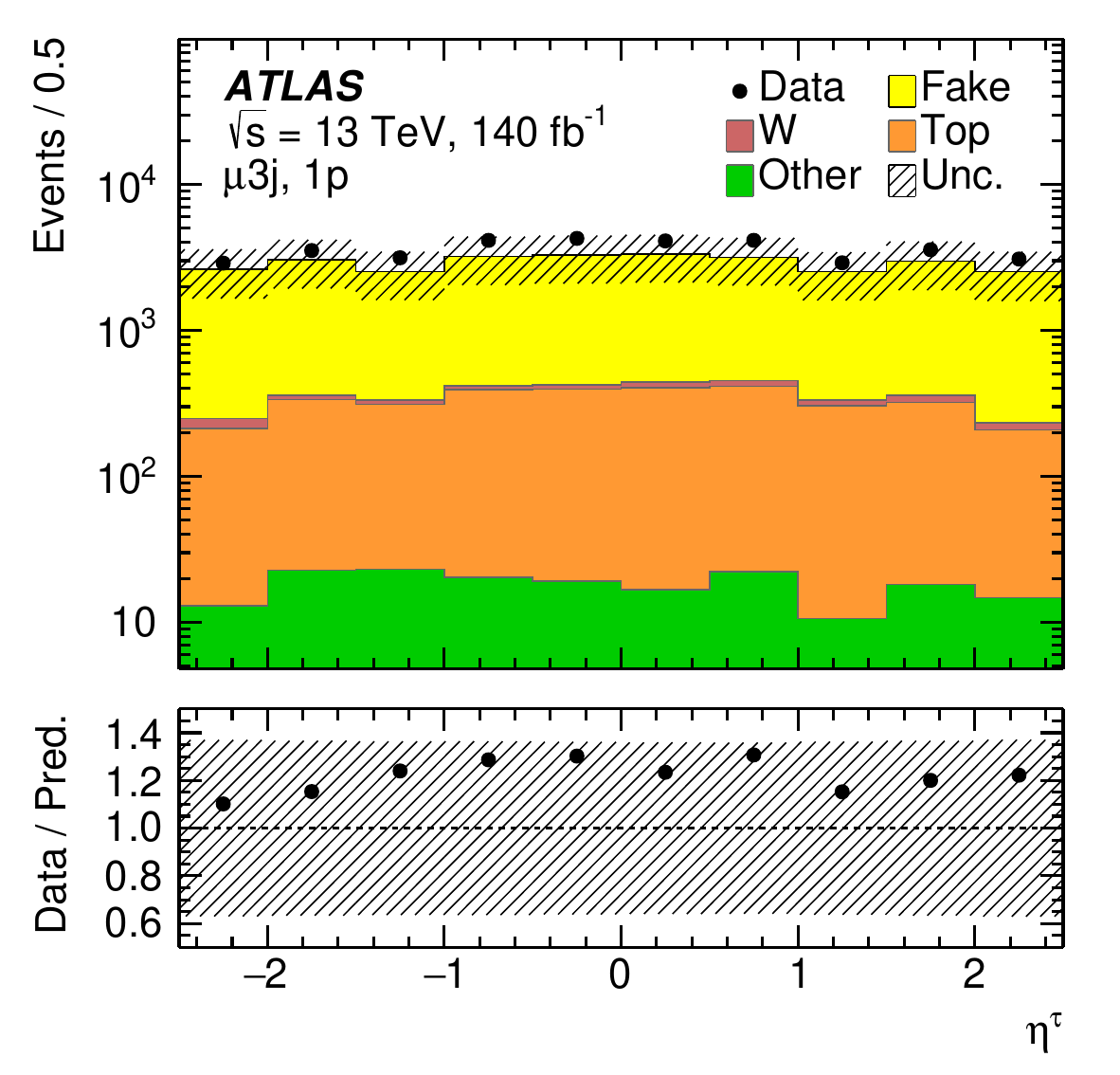}
\label{fig:spectra_mu3j_1p_tau0_eta}
}\quad
\subfloat[]{
\includegraphics[width=0.32\textwidth]{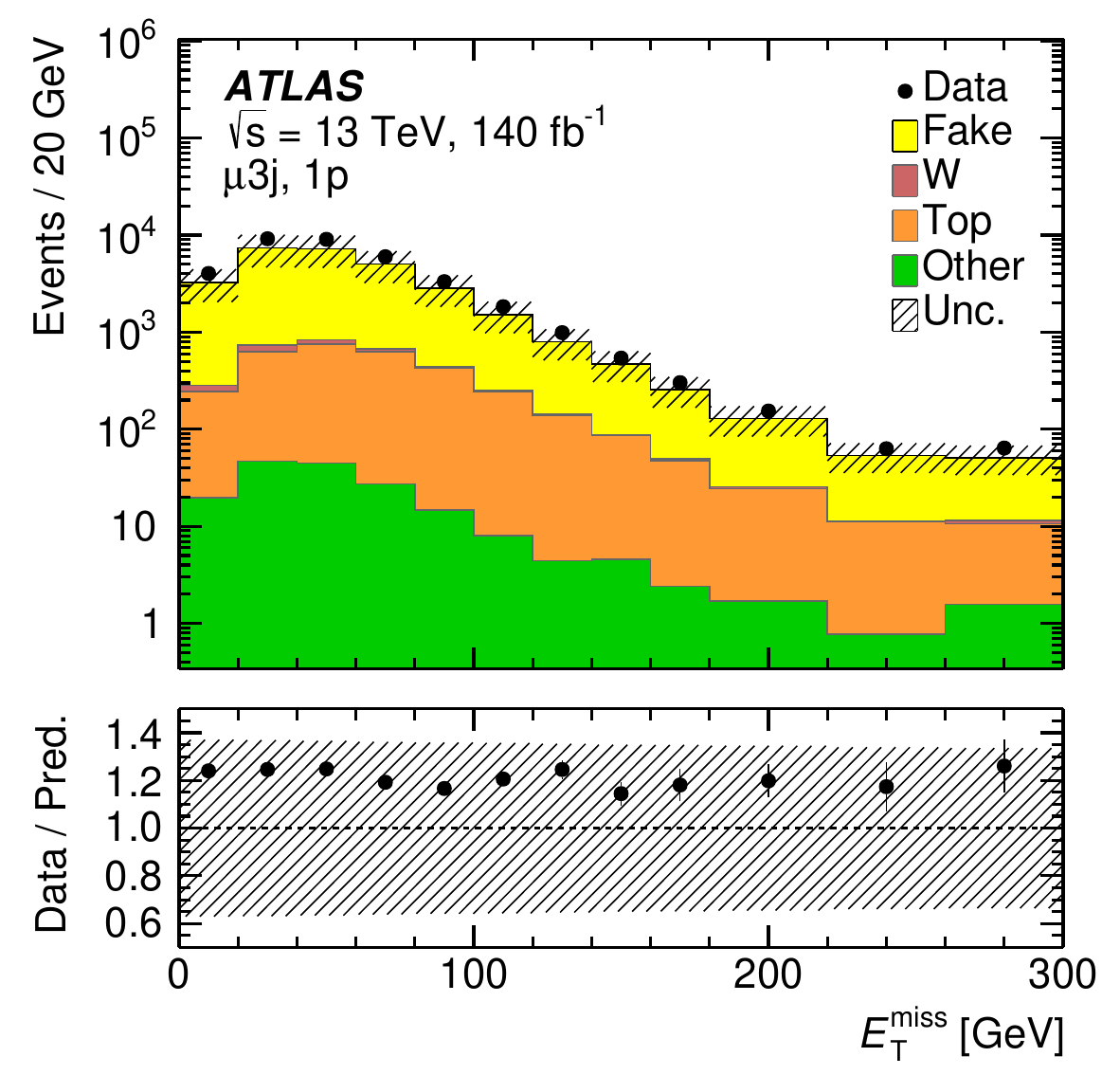}
\label{fig:spectra_mu3j_1p_met}
}\quad
\subfloat[]{
\includegraphics[width=0.32\textwidth]{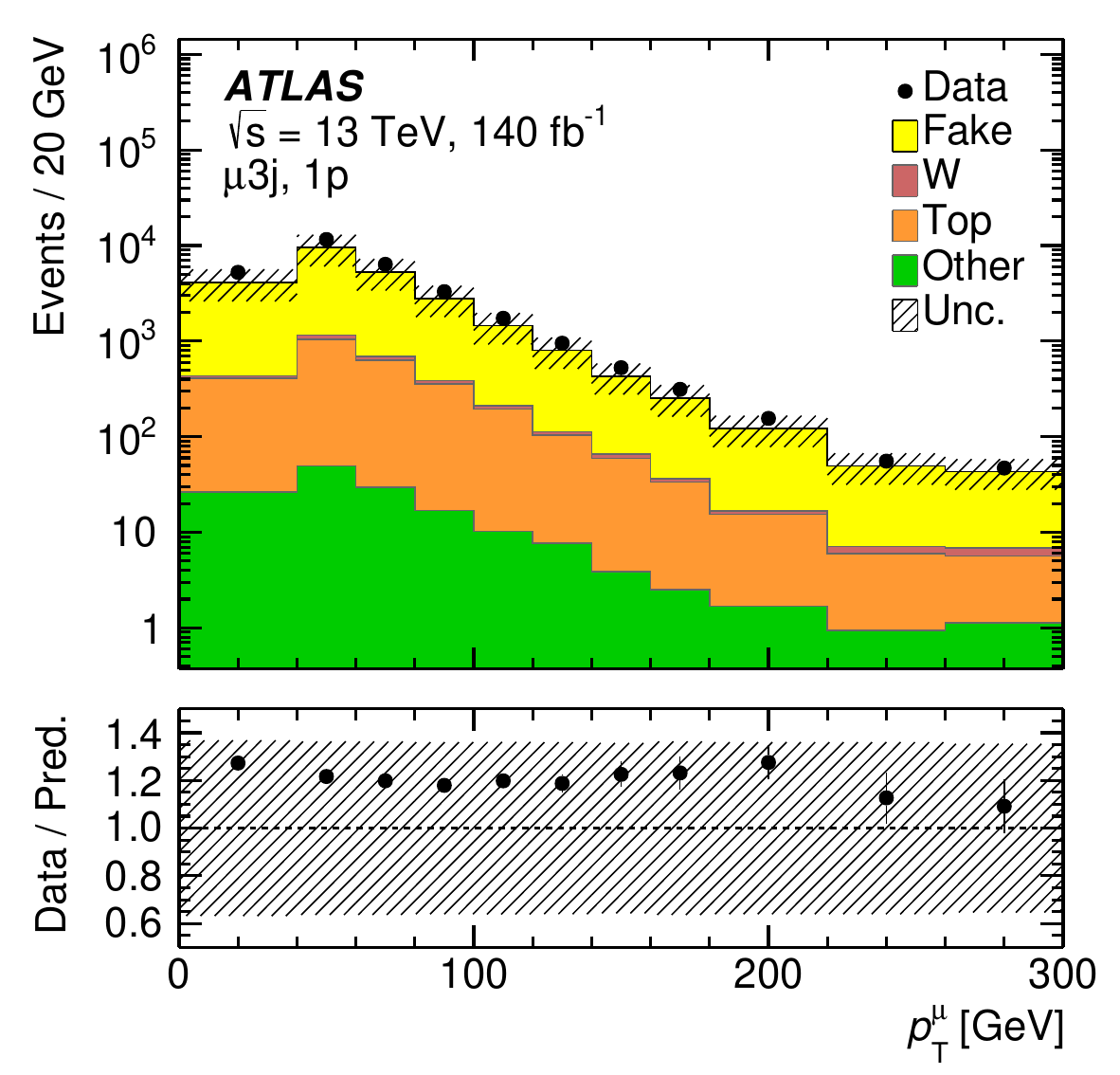}
\label{fig:spectra_mu3j_1p_mu0_pt}
}\quad
\subfloat[]{
\includegraphics[width=0.32\textwidth]{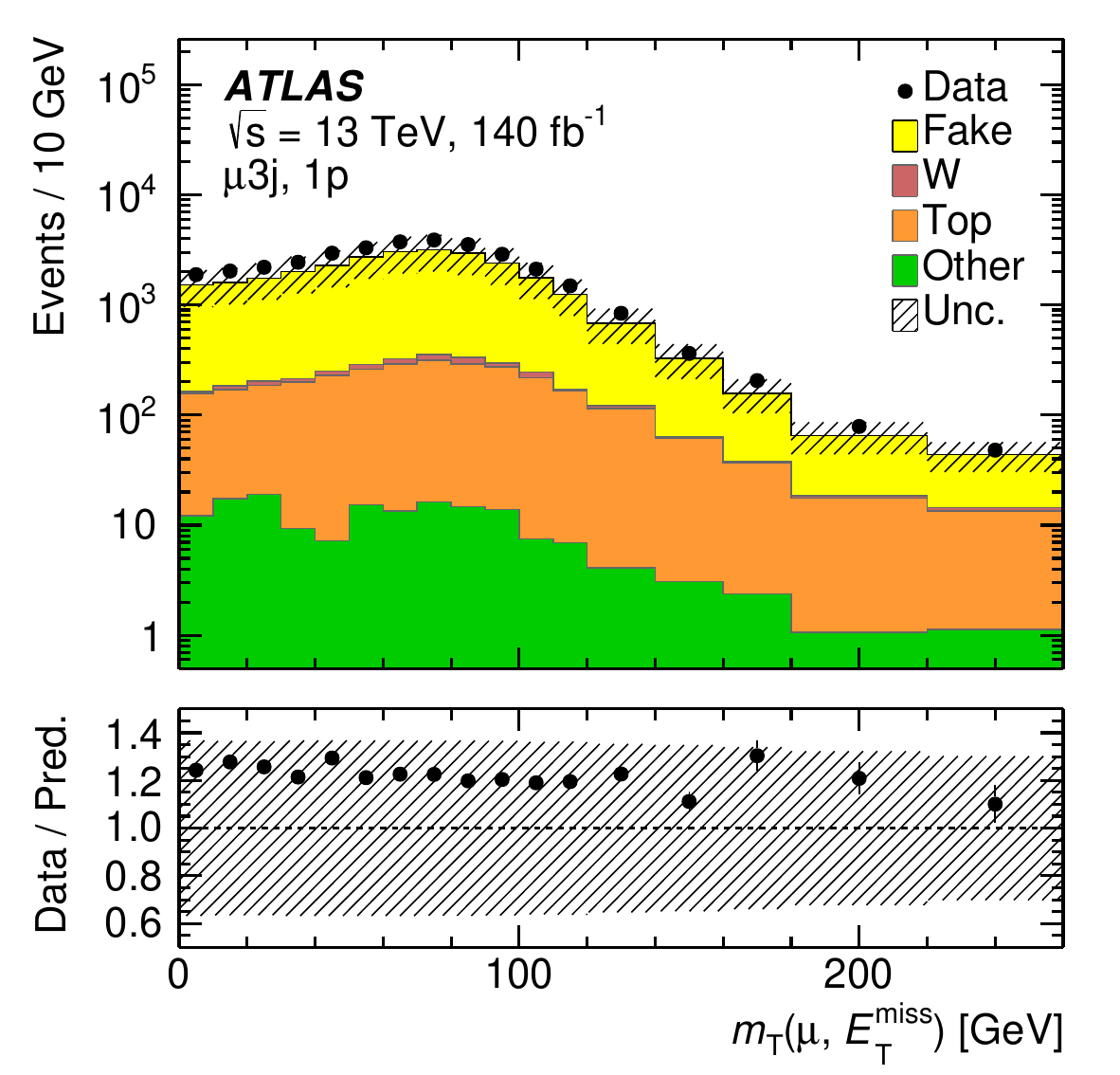}
\label{fig:spectra_mu3j_1p_mu_met_mt}
}\quad
\subfloat[]{
\includegraphics[width=0.32\textwidth]{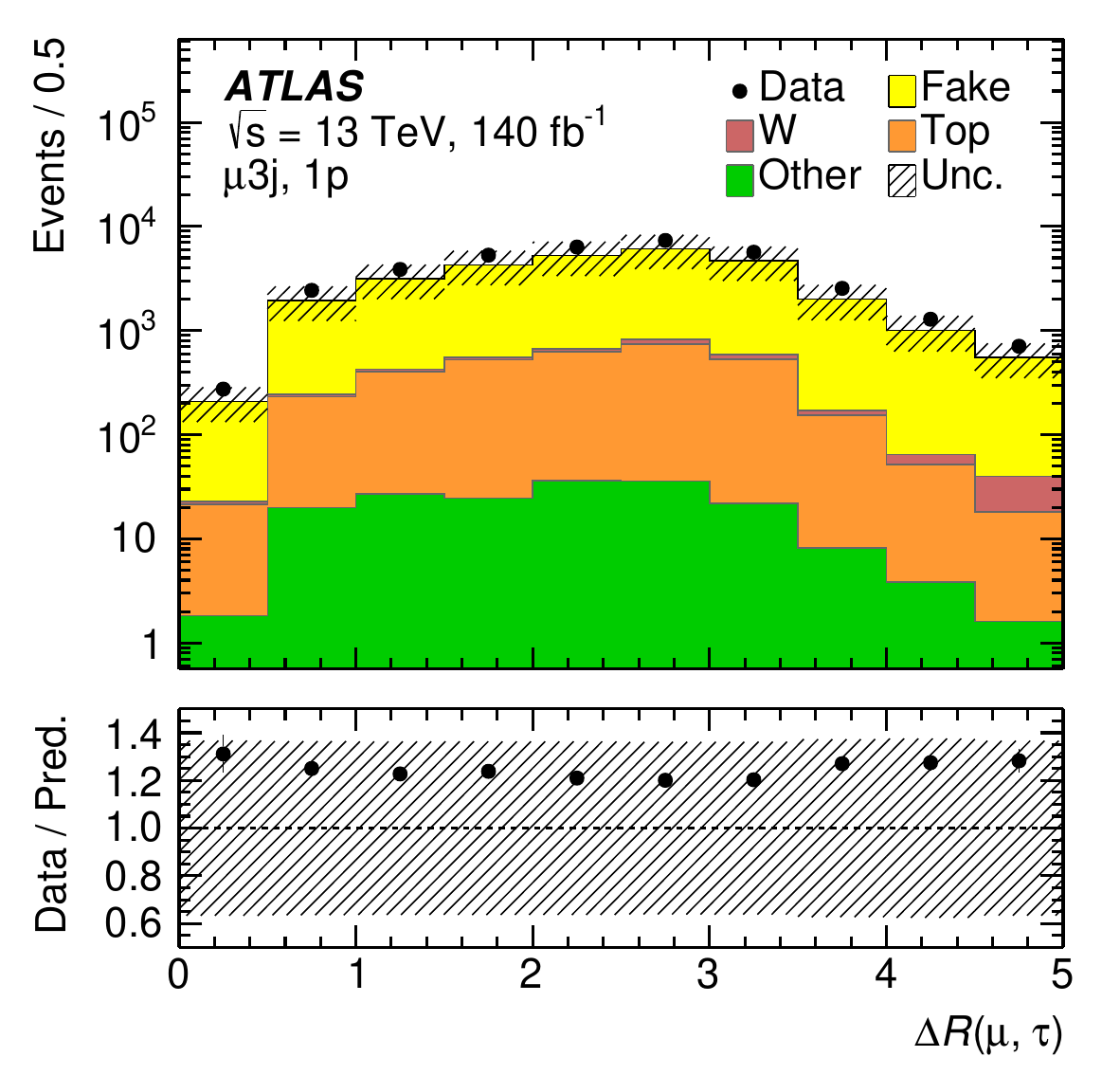}
\label{fig:spectra_mu3j_1p_mu_tau_dR}
}
\caption{
Comparison of the 1-prong (1p) fake-background prediction (yellow area) with the data (black points) in the \muthreej region.
The real-\tl-lepton backgrounds are also shown.
The shaded band displays the total uncertainty of the prediction.
It includes the fake-related systematic uncertainty, the statistical uncertainty of the \ffuff,
and the uncertainty due to the limited number of events in the antiID subregion of the SR and the MC samples.
The comparison is shown for several variables: (a) \tauhadvis transverse momentum, (b) \tauhadvis $\eta$, (c) \met, (d) \ptmu, (e) \mumetmt, and (f) \mutaudr.
The compact notation \pttau and \etatau is used in the $x$-axis labels of figures (a) and (b) to
denote the transverse momentum and $\eta$ of \tauhadvis.
}
\label{fig:spectra_mu3j_1p}
\end{figure}

\begin{figure}
\centering
\subfloat[]{
\includegraphics[width=0.32\textwidth]{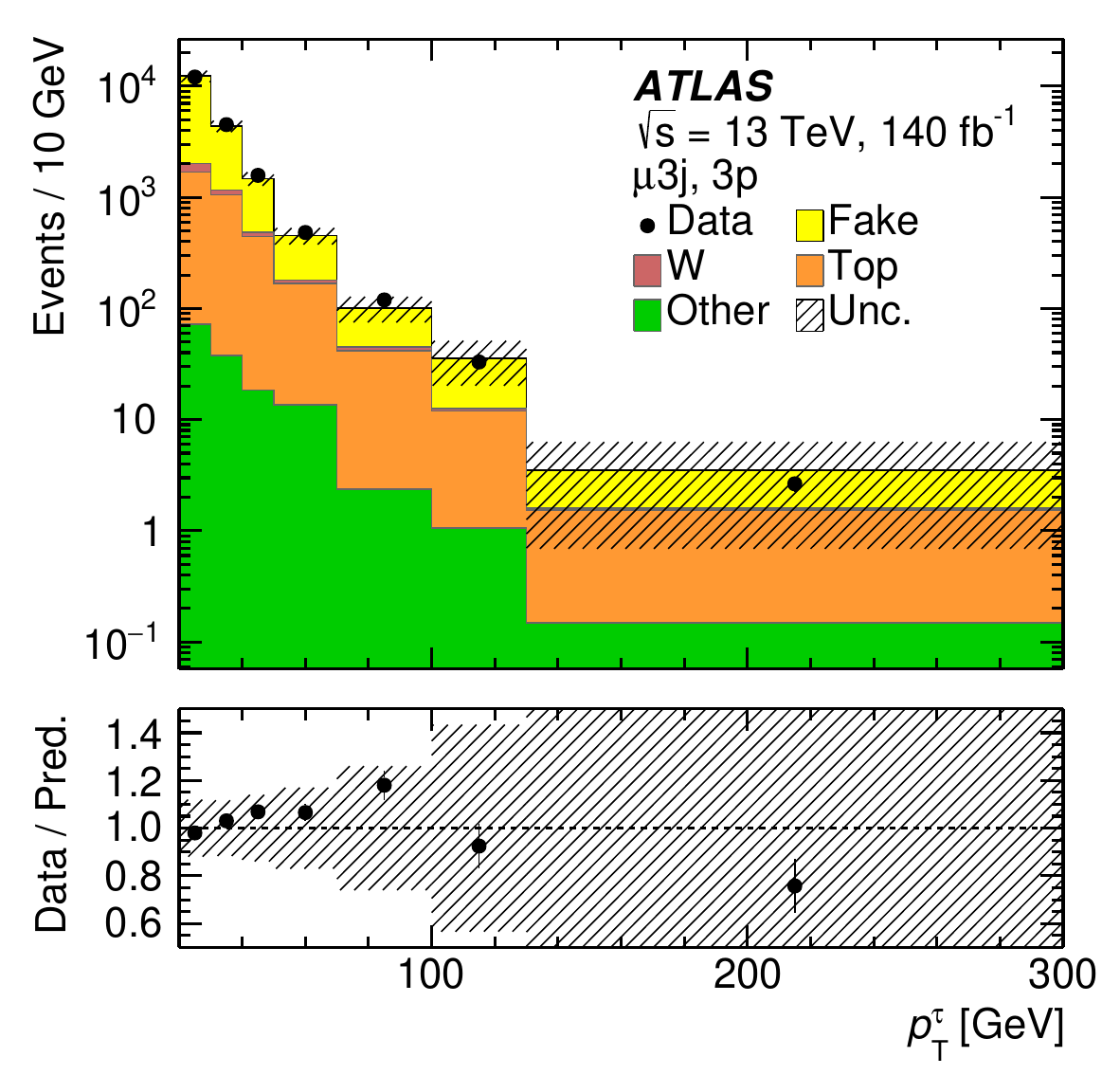}
\label{fig:spectra_mu3j_3p_tau0_pt}
}\quad
\subfloat[]{
\includegraphics[width=0.32\textwidth]{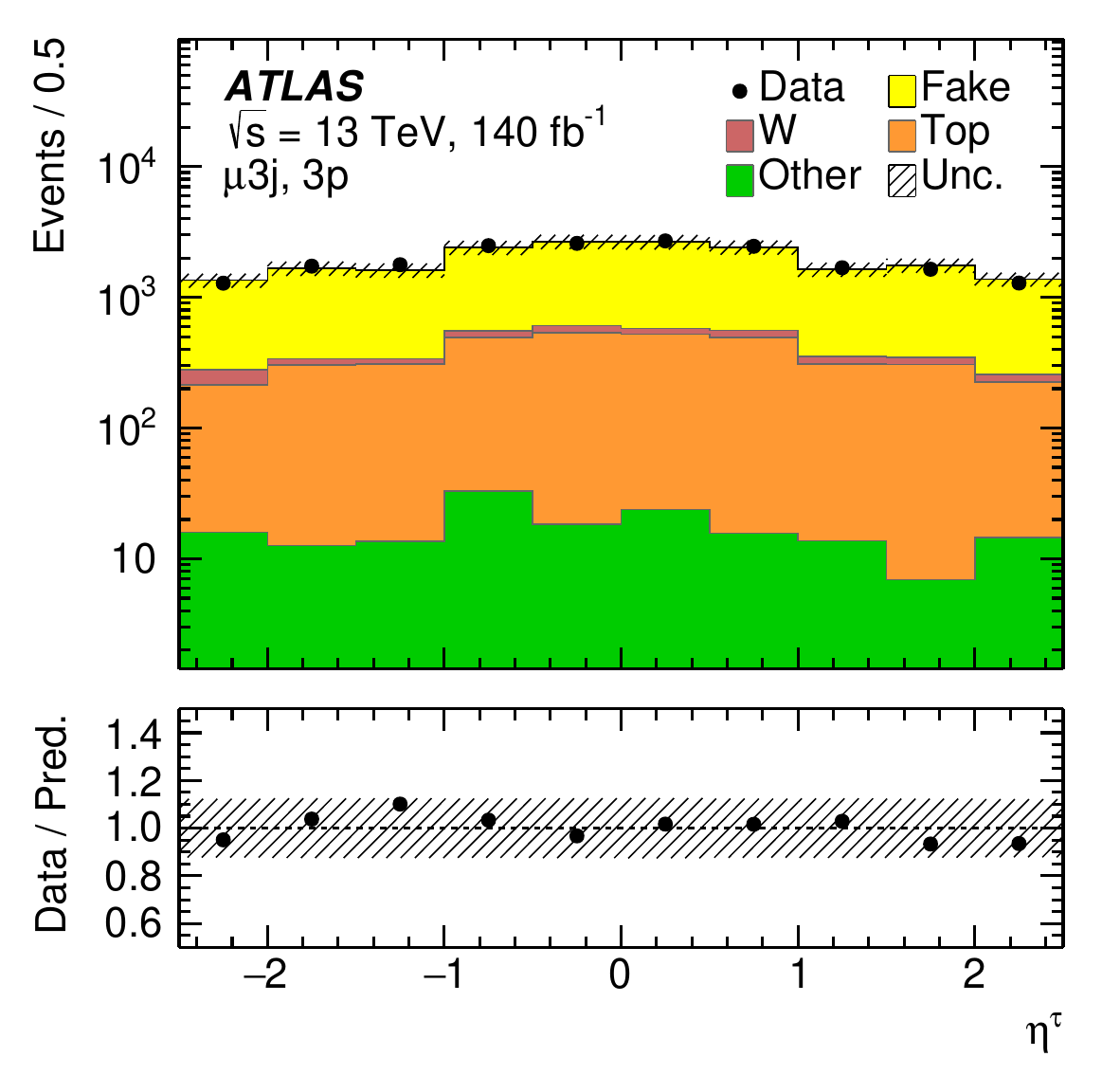}
\label{fig:spectra_mu3j_3p_tau0_eta}
}\quad
\subfloat[]{
\includegraphics[width=0.32\textwidth]{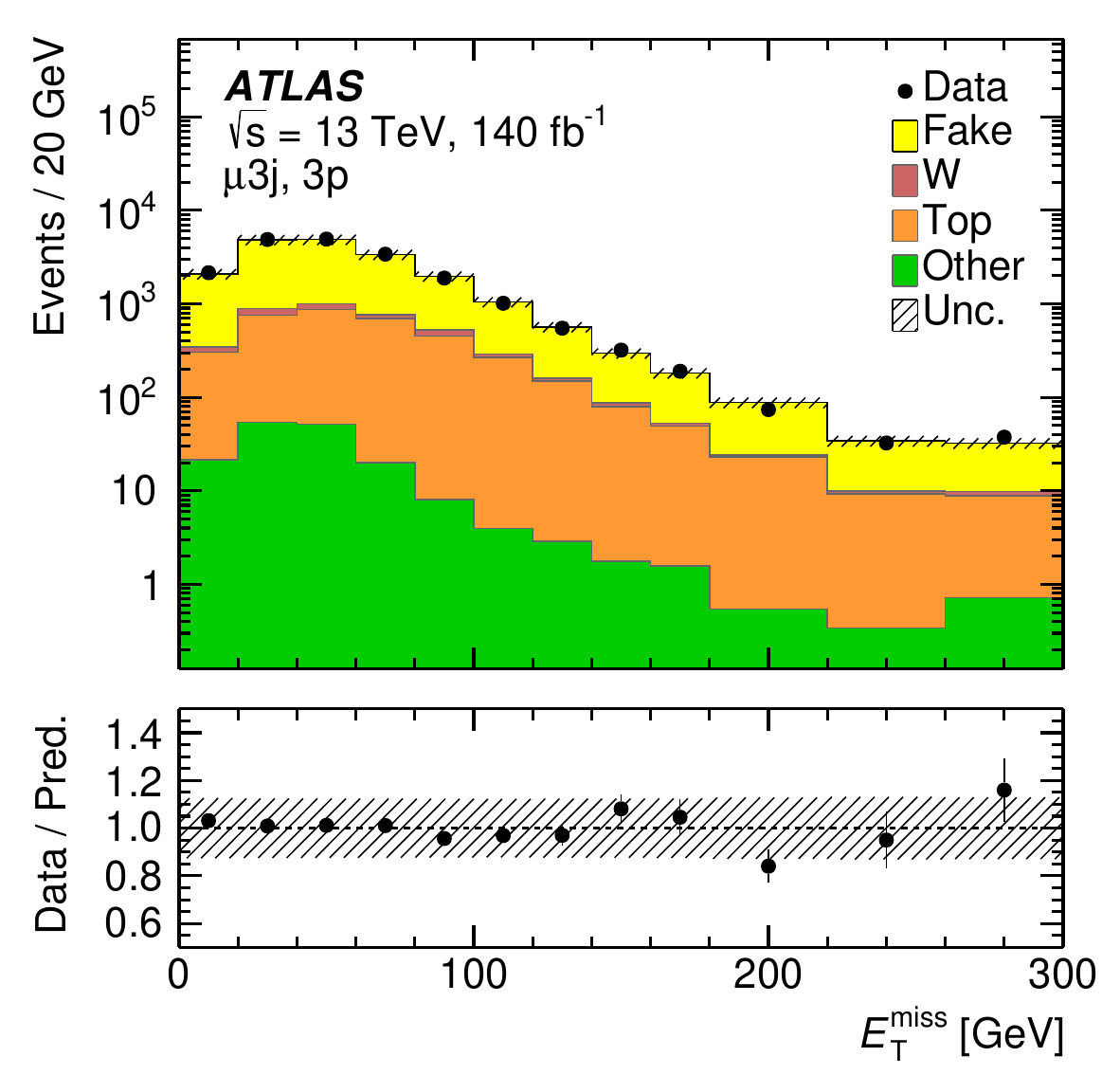}
\label{fig:spectra_mu3j_3p_met}
}\quad
\subfloat[]{
\includegraphics[width=0.32\textwidth]{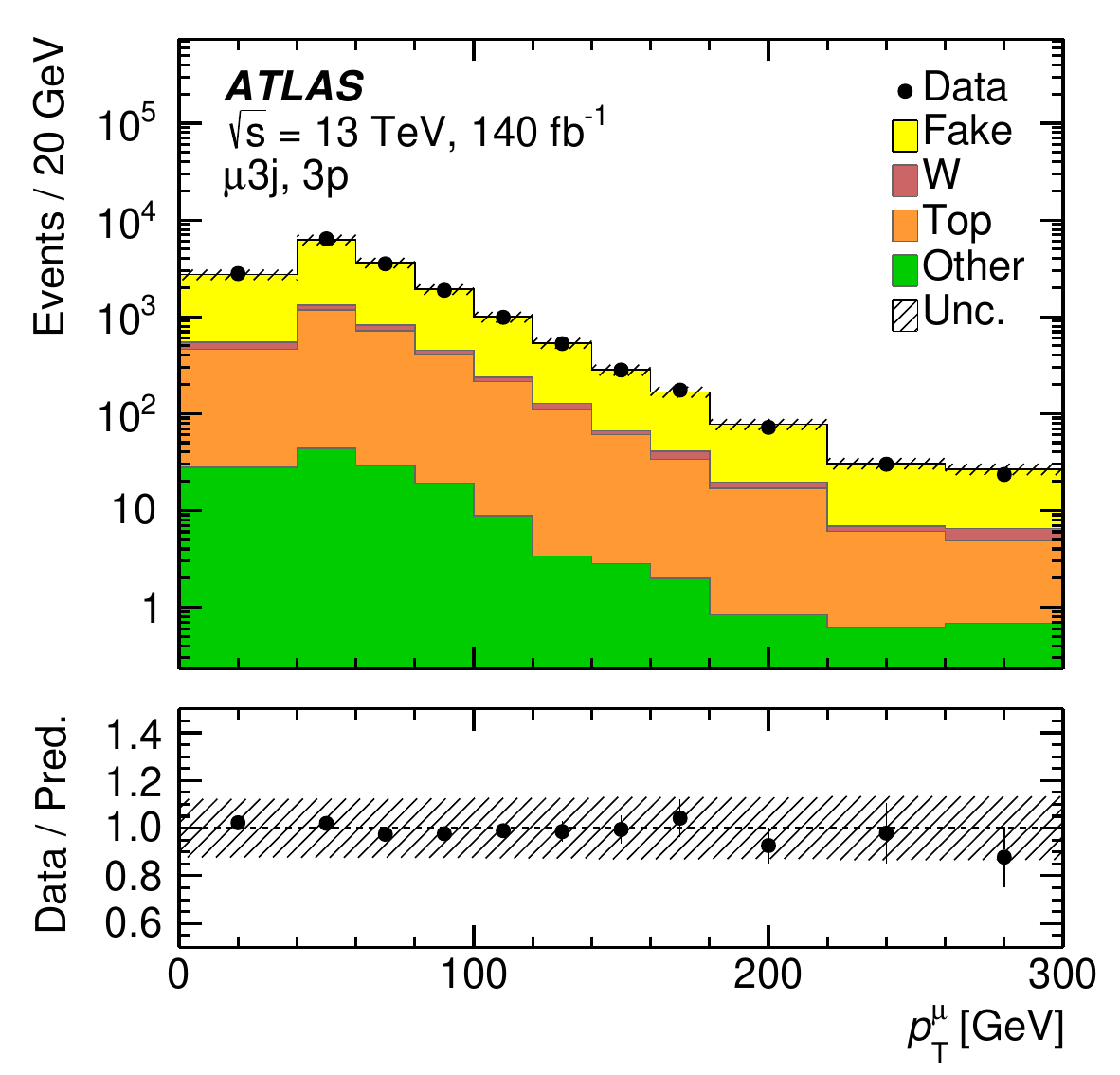}
\label{fig:spectra_mu3j_3p_mu0_pt}
}\quad
\subfloat[]{
\includegraphics[width=0.32\textwidth]{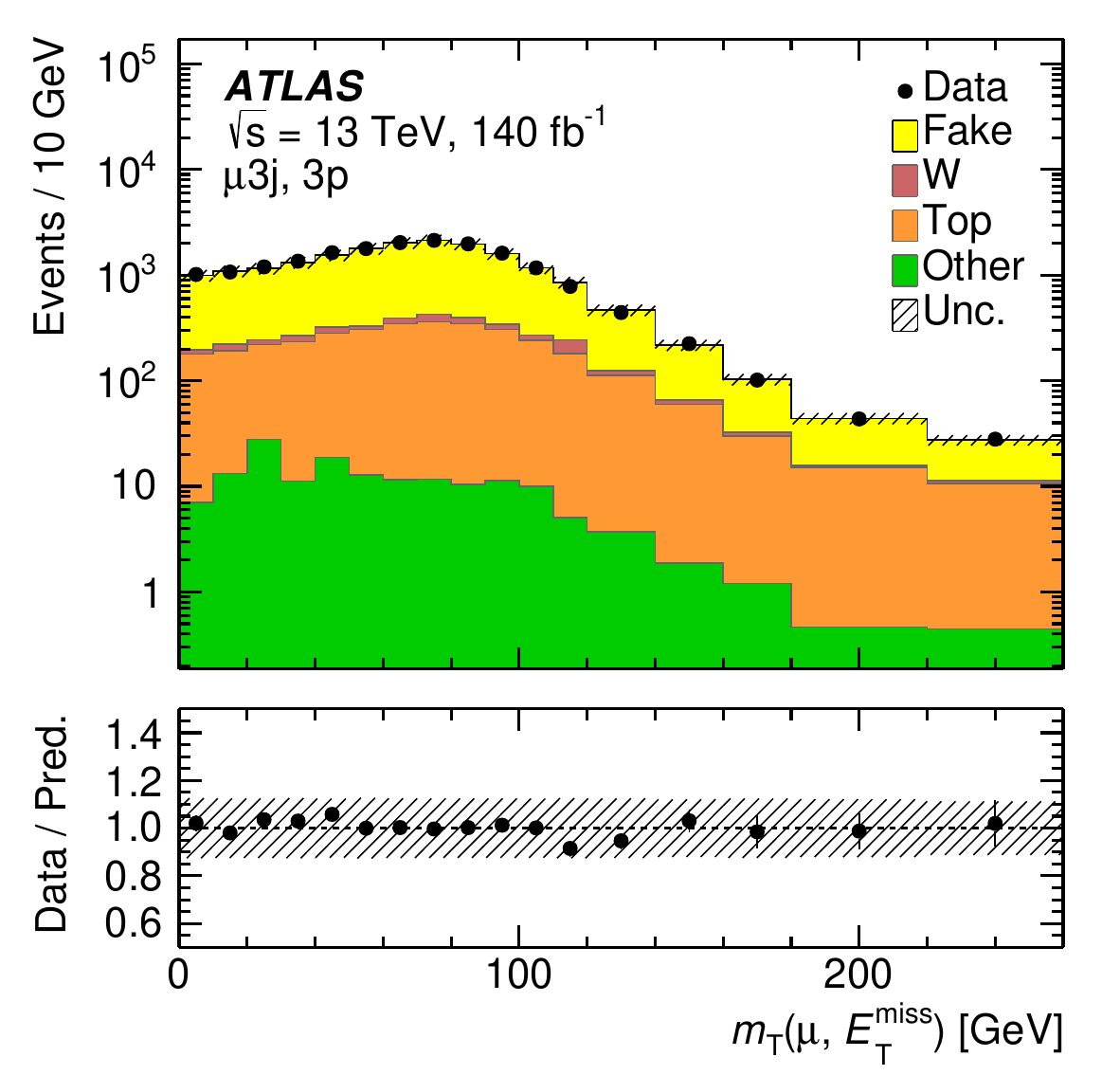}
\label{fig:spectra_mu3j_3p_mu_met_mt}
}\quad
\subfloat[]{
\includegraphics[width=0.32\textwidth]{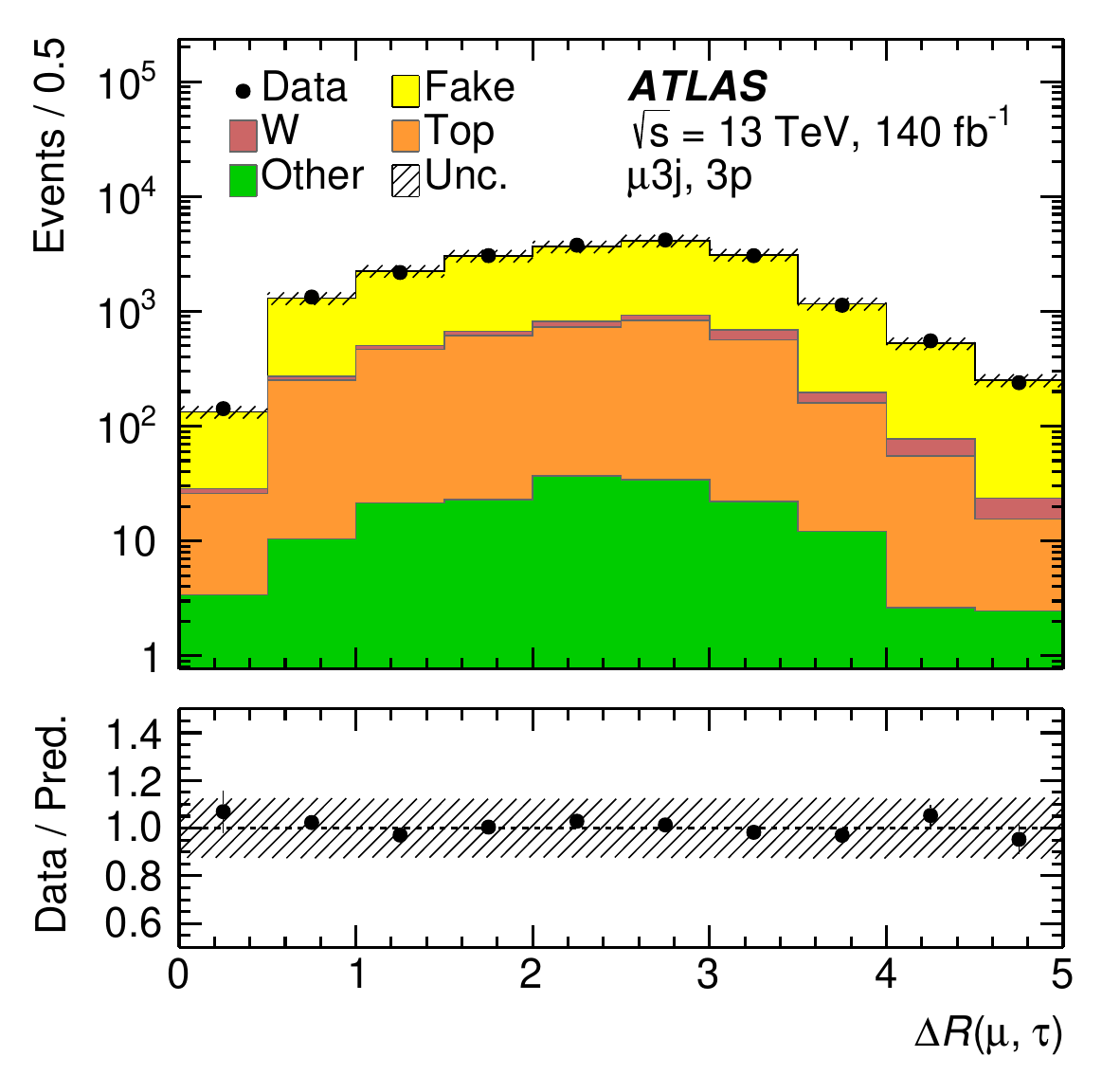}
\label{fig:spectra_mu3j_3p_mu_tau_dR}
}
\caption{
Comparison of the 3-prong (3p) fake-background prediction (yellow area) with the data (black points) in the \muthreej region.
The real-\tl-lepton backgrounds are also shown.
The shaded band displays the total uncertainty of the prediction.
It includes the fake-related systematic uncertainty, the statistical uncertainty on the \ffuff,
and the uncertainty due to the limited number of events in the antiID subregion of the SR and the MC samples.
The comparison is shown for several variables: (a) \tauhadvis transverse momentum, (b) \tauhadvis $\eta$, (c) \met, (d) \ptmu, (e) \mumetmt, and (f) \mutaudr.
The compact notation \pttau and \etatau is used in the $x$-axis labels of Figures (a) and (b) to
denote the transverse momentum and $\eta$ of \tauhadvis.
}
\label{fig:spectra_mu3j_3p}
\end{figure}

Preliminary versions of the UFF method, using a lower number of DRs, a different discriminating variable and/or different systematic uncertainty estimates, were successfully used for
fake-background estimation in three published ATLAS analyses.
Final states $\tauhadvis\ +$ jets, $\tauhadvis+e/\mu$, $\tauhadvis+\met$, $2\tauhadvis+\met/b$-jets, and $2\tauhadvis$ inclusive have been probed.


%

%

\section{Applicability and limitations of the Universal Fake Factor method}
\label{sec:proscons}

The UFF method offers a valuable approach for estimating fake-\tauhadvis background.
However, it is crucial to acknowledge its advantages and limitations to delineate the types of analyses for which it is best suited.
The UFF method resolves the problem related to the a priori unknown fake-\tauhadvis composition of a SR.
The FFs suitable for a SR are evaluated as a linear combination of FFs determined in four predefined DRs, \Zmm, \ttbar, \mjh, and \mjl.
The related systematic uncertainty is physically motivated and simple to evaluate.
The only input required to use the UFF method in an analysis are the q/g tagger score distributions (data and the MC-based real-\tl-lepton contamination estimate) in the different phase space bins of the antiID subregion of the SR.
The analysis using the UFF method does not need to exploit regions of phase space similar to the SR (in terms of the selection criteria) to determine FFs suitable for the analysis' SR.
These regions can be employed for a detailed validation of the UFF-method-based prediction.

The UFF method is best suited for analyses with fake-\tauhadvis with \pt below $\sim150~\gev$ and a statistical uncertainty of the background prediction larger than the UFF-related systematic uncertainty.
Such analyses can highly benefit from the predefined UFF methodology and the ready-to-use DR FFs and templates, resulting in faster analysis development.

The applicability of the UFF method to analyses involving very high-\pt \tauhadvis warrants examination.
In the present paper, the method is tested for \tauhadvis \pt of up to $\sim200-300~\gev$, see figures~\ref{fig:spectra_wmunu_1p_tau0_pt}, \ref{fig:spectra_wmunu_3p_tau0_pt}, \ref{fig:spectra_mu3j_1p_tau0_pt}, and \ref{fig:spectra_mu3j_3p_tau0_pt}.
The \pt reach is mainly limited by the available statistics in the DRs.
In Figure~\ref{fig:uff_ff}, the highest-\pt bin takes into account the overflow events containing \tauhadvis candidates with \pt up to $300~\gev$, but the \pt binning can be adjusted to accommodate the analysis needs.
When deciding to use fine binning in the high-\pt region, the analysis using the UFF method should consider assigning a related systematic uncertainty.

A second question the analysis using the UFF method should address is the size of the total systematic uncertainty.
Precision measurements and systematics-dominated searches might prefer to use the original FF method and a set of dedicated DRs defined by selection criteria similar to those in the SR if the DRs have high enough statistics
and the extrapolation of the FFs to the SR has lower systematic uncertainty.
The decision has to be made on a case-by-case basis, and no general rule can be given.

A problem arises if the SR largely overlaps with the DRs.
In this case, the UFF method is not applicable, as the FFs in the DRs are not independent of the SR.
The fraction of overlapping events must be checked before making the decision to use the UFF method.

Analyses dominated by \qfake fakes and sensitive to the amplitude of the uncertainty band can mitigate potential bias by deriving the UFF templates and FFs from \wmunu events instead of \Zmm.
This substitution can shift the nominal prediction by about $1\sigma$ of the quoted systematic uncertainty and provide an alternative modelling with different kinematic composition.
The analysis can take the envelope of the $Z$- and $W$-based implementations as an additional shape uncertainty, validate with closure tests in 3-prong-enhanced control regions,
and profile the corresponding nuisance parameters independently in the fit to reduce fit-induced artifacts in high-mass searches.

Analyses dominated by rare fake-\tauhadvis sources, where the physics origin of the fakes induces correlations not fully captured by the envelope method, might require a refinement of the systematic uncertainty estimation described in Section~\ref{sec:uff_unc}. For example, a hypothetical analysis dominated by \bfake fakes from $\Zmm + b$-jet events would be sensitive to the sign of the difference between the FFs derived from \ttbar and $\Zmm + b$ samples.
In such a case, it would be more appropriate to use the relative difference between the two FF sets as the \ttbar systematic uncertainty, taking the sign of the difference into account.
For analyses where the MC-based prediction may lie outside the envelope defined by the reference regions, making the symmetrization both impactful and potentially inappropriate,  a more tailored treatment should be adopted -- for example, by retaining the sign of the difference or by constructing an analysis-specific uncertainty model that preserves genuine shape dependencies.

This paper presents the UFF method using the \textit{Loose} \tauhadvis ID working point,
but the method can be adapted to other \tauhadvis ID selections.
Different inputs to the UFF method are required for each \tauhadvis ID efficiency working point.
Similarly, recalculations are required for analyses utilizing \tauhadvis triggers
that include specific \tauhadvis ID criteria.
Finally, if the q/g tagger is updated, its templates will also need to be recalculated in the antiID
\Zmm, \ttbar, \mjh and \mjl subregions.


%

%
%
\FloatBarrier

%

\section{Conclusions}
\label{sec:conclusion}

This paper presents the UFF method, a generalization of the FF method, commonly used by the ATLAS experiment at the LHC to estimate the fake-\tauhadvis background.
The UFF method provides a prescription for estimating the FFs suitable for the SR of an analysis.
The method does not require to define one or more dedicated DRs
for the FFs measurement; instead, it is based on the usage of four predefined regions
enriched in different types of fakes.
The \Zmm, \ttbar, \mjh and \mjl regions target \qfake fakes (from light-quark jets), \bfake fakes (from \bjets),
\gfake fakes (from gluon-initiated jets), and \pfake fakes (from \pileup jets), respectively.
The analysis is provided with the FFs and q/g tagger score spectra determined in the corresponding antiID subregions.
The only input the analysis has to prepare is the set of q/g tagger score spectra corresponding to the various \pt and prongness bins of its antiID subregion of the SR.
The UFF method then estimates the suitable FFs using a template fit in each \pt and prongness bin and a simple formula
for the combination of the four provided FFs.
The method was successfully applied in several phase-space regions.
In particular, the \wmunu region shows that the UFF method can predict the FFs and the fake-\tauhadvis background within the associated systematic uncertainty.
The present study used the full \RunTwo \pp collision data set, and the systematic uncertainty of the FFs for the \wmunu region
was estimated to be between 15\% and 35\%, depending on the fake-\tauhadvis \pt and prongness.
A testing procedure applied to the \muthreej region demonstrated that the UFF method also works well in a region enriched in \bfake fakes.

Besides providing a simple generic algorithm and the necessary inputs, the UFF method has one more advantage.
The fake-background prediction obtained with the UFF method can be validated in the DRs that the original FF method would use for
the FFs measurement and the estimation of systematic uncertainties.
This resolves a serious drawback of the original FF method, that it is very demanding in terms of the number of
required DRs, and an exhaustive validation of the predictions often is not possible.
It is also important to emphasize that the UFF method is data-driven to a large extent.
MC simulation is used only for the subtraction of the real-\tl-lepton contamination and the systematic uncertainty estimation.
The UFF method is thus a powerful tool that can be used in a wide range of analyses.


%

%
%
%
%
%
\section*{Acknowledgements}
%

%

%
%

%
%

We thank CERN for the very successful operation of the LHC and its injectors, as well as the support staff at
CERN and at our institutions worldwide without whom ATLAS could not be operated efficiently.

The crucial computing support from all WLCG partners is acknowledged gratefully, in particular from CERN, the ATLAS Tier-1 facilities at TRIUMF/SFU (Canada), NDGF (Denmark, Norway, Sweden), CC-IN2P3 (France), KIT/GridKA (Germany), INFN-CNAF (Italy), NL-T1 (Netherlands), PIC (Spain), RAL (UK) and BNL (USA), the Tier-2 facilities worldwide and large non-WLCG resource providers. Major contributors of computing resources are listed in Ref.~\cite{ATL-SOFT-PUB-2025-001}.

We gratefully acknowledge the support of ANPCyT, Argentina; YerPhI, Armenia; ARC, Australia; BMWFW and FWF, Austria; ANAS, Azerbaijan; CNPq and FAPESP, Brazil; NSERC, NRC and CFI, Canada; CERN; ANID, Chile; CAS, MOST and NSFC, China; Minciencias, Colombia; MEYS CR, Czech Republic; DNRF and DNSRC, Denmark; IN2P3-CNRS and CEA-DRF/IRFU, France; SRNSFG, Georgia; BMFTR, HGF and MPG, Germany; GSRI, Greece; RGC and Hong Kong SAR, China; ICHEP and Academy of Sciences and Humanities, Israel; INFN, Italy; MEXT and JSPS, Japan; CNRST, Morocco; NWO, Netherlands; RCN, Norway; MNiSW, Poland; FCT, Portugal; MNE/IFA, Romania; MSTDI, Serbia; MSSR, Slovakia; ARIS and MVZI, Slovenia; DSI/NRF, South Africa; MICIU/AEI, Spain; SRC and Wallenberg Foundation, Sweden; SERI, SNSF and Cantons of Bern and Geneva, Switzerland; NSTC, Taipei; TENMAK, T\"urkiye; STFC/UKRI, United Kingdom; DOE and NSF, United States of America.

Individual groups and members have received support from BCKDF, CANARIE, CRC and DRAC, Canada; CERN-CZ, FORTE and PRIMUS, Czech Republic; COST, ERC, ERDF, Horizon 2020, ICSC-NextGenerationEU and Marie Sk{\l}odowska-Curie Actions, European Union; Investissements d'Avenir Labex, Investissements d'Avenir Idex and ANR, France; DFG and AvH Foundation, Germany; Herakleitos, Thales and Aristeia programmes co-financed by EU-ESF and the Greek NSRF, Greece; BSF-NSF and MINERVA, Israel; NCN and NAWA, Poland; La Caixa Banking Foundation, CERCA Programme Generalitat de Catalunya and PROMETEO and GenT Programmes Generalitat Valenciana, Spain; G\"{o}ran Gustafssons Stiftelse, Sweden; The Royal Society and Leverhulme Trust, United Kingdom.

In addition, individual members wish to acknowledge support from Armenia: Yerevan Physics Institute (FAPERJ); CERN: European Organization for Nuclear Research (CERN DOCT); Chile: Agencia Nacional de Investigaci\'on y Desarrollo (FONDECYT 1230812, FONDECYT 1240864, Fondecyt 3240661); China: Chinese Ministry of Science and Technology (MOST-2023YFA1605700, MOST-2023YFA1609300), National Natural Science Foundation of China (NSFC - 12175119, NSFC 12275265, NSFC-12075060); Czech Republic: Czech Science Foundation (GACR - 24-11373S), Ministry of Education Youth and Sports (ERC-CZ-LL2327, FORTE CZ.02.01.01/00/22\_008/0004632), PRIMUS Research Programme (PRIMUS/21/SCI/017); EU: H2020 European Research Council (ERC - 101002463); European Union: European Research Council (BARD No. 101116429, ERC - 948254, ERC 101089007), Horizon 2020 Framework Programme (MUCCA - CHIST-ERA-19-XAI-00), European Union, Future Artificial Intelligence Research (FAIR-NextGenerationEU PE00000013), Italian Center for High Performance Computing, Big Data and Quantum Computing (ICSC, NextGenerationEU); France: Agence Nationale de la Recherche (ANR-20-CE31-0013, ANR-21-CE31-0013, ANR-21-CE31-0022, ANR-22-EDIR-0002); Germany: Baden-Württemberg Stiftung (BW Stiftung-Postdoc Eliteprogramme), Deutsche Forschungsgemeinschaft (DFG - 469666862, DFG - CR 312/5-2); Italy: Istituto Nazionale di Fisica Nucleare (ICSC, NextGenerationEU), Ministero dell'Università e della Ricerca (NextGenEU 153D23001490006 M4C2.1.1, NextGenEU I53D23000820006 M4C2.1.1, NextGenEU I53D23001490006 M4C2.1.1); Japan: Japan Society for the Promotion of Science (JSPS KAKENHI JP22H01227, JSPS KAKENHI JP22H04944, JSPS KAKENHI JP22KK0227, JSPS KAKENHI JP23KK0245); Norway: Research Council of Norway (RCN-314472); Poland: Ministry of Science and Higher Education (IDUB AGH, POB8, D4 no 9722), Polish National Agency for Academic Exchange (PPN/PPO/2020/1/00002/U/00001), Polish National Science Centre (NCN 2021/42/E/ST2/00350, NCN OPUS 2023/51/B/ST2/02507, NCN OPUS nr 2022/47/B/ST2/03059, NCN UMO-2019/34/E/ST2/00393, UMO-2020/37/B/ST2/01043, UMO-2021/40/C/ST2/00187, UMO-2022/47/O/ST2/00148, UMO-2023/49/B/ST2/04085, UMO-2023/51/B/ST2/00920); Portugal: Foundation for Science and Technology (FCT); Spain: Generalitat Valenciana (Artemisa, FEDER, IDIFEDER/2018/048), Ministry of Science and Innovation (MCIN \& NextGenEU PCI2022-135018-2, MICIN \& FEDER PID2021-125273NB, RYC2019-028510-I, RYC2020-030254-I, RYC2021-031273-I, RYC2022-038164-I); Sweden: Carl Trygger Foundation (Carl Trygger Foundation CTS 22:2312), Swedish Research Council (Swedish Research Council 2023-04654, VR 2018-00482, VR 2021-03651, VR 2022-03845, VR 2022-04683, VR 2023-03403), Knut and Alice Wallenberg Foundation (KAW 2018.0458, KAW 2019.0447, KAW 2022.0358); Switzerland: Swiss National Science Foundation (SNSF - PCEFP2\_194658); United Kingdom: Leverhulme Trust (Leverhulme Trust RPG-2020-004), Royal Society (NIF-R1-231091); United States of America: U.S. Department of Energy (ECA DE-AC02-76SF00515), Neubauer Family Foundation.

%
%


%
%
%
%
%
%
%
%
%

%
%
%
\printbibliography
\clearpage
 
\begin{flushleft}
\hypersetup{urlcolor=black}
{\Large The ATLAS Collaboration}

\bigskip

\AtlasOrcid[0000-0002-6665-4934]{G.~Aad}$^\textrm{\scriptsize 105}$,
\AtlasOrcid[0000-0001-7616-1554]{E.~Aakvaag}$^\textrm{\scriptsize 17}$,
\AtlasOrcid[0000-0002-5888-2734]{B.~Abbott}$^\textrm{\scriptsize 124}$,
\AtlasOrcid[0000-0002-0287-5869]{S.~Abdelhameed}$^\textrm{\scriptsize 120a}$,
\AtlasOrcid[0000-0002-1002-1652]{K.~Abeling}$^\textrm{\scriptsize 56}$,
\AtlasOrcid[0000-0001-5763-2760]{N.J.~Abicht}$^\textrm{\scriptsize 50}$,
\AtlasOrcid[0000-0002-8496-9294]{S.H.~Abidi}$^\textrm{\scriptsize 30}$,
\AtlasOrcid[0009-0003-6578-220X]{M.~Aboelela}$^\textrm{\scriptsize 46}$,
\AtlasOrcid[0000-0002-9987-2292]{A.~Aboulhorma}$^\textrm{\scriptsize 36e}$,
\AtlasOrcid[0000-0001-5329-6640]{H.~Abramowicz}$^\textrm{\scriptsize 158}$
% ...
\bigskip
\\

% ...
$^{17}$Department for Physics and Technology, University of Bergen, Bergen; Norway.\\
% ...
$^{30}$Physics Department, Brookhaven National Laboratory, Upton NY; United States of America.\\
% ...
$^{36}$$^{(a)}$Facult\'e des Sciences Ain Chock, Universit\'e Hassan II de Casablanca;$^{(b)}$Facult\'{e} des Sciences, Universit\'{e} Ibn-Tofail, K\'{e}nitra;$^{(c)}$Facult\'e des Sciences Semlalia, Universit\'e Cadi Ayyad, LPHEA-Marrakech;$^{(d)}$LPMR, Facult\'e des Sciences, Universit\'e Mohamed Premier, Oujda;$^{(e)}$Facult\'e des sciences, Universit\'e Mohammed V, Rabat;$^{(f)}$Institute of Applied Physics, Mohammed VI Polytechnic University, Ben Guerir; Morocco.\\
% ...
$^{46}$Physics Department, Southern Methodist University, Dallas TX; United States of America.\\
% ...
$^{50}$Fakult\"{a}t Physik , Technische Universit{\"a}t Dortmund, Dortmund; Germany.\\
% ...
$^{56}$II. Physikalisches Institut, Georg-August-Universit\"{a}t G\"ottingen, G\"ottingen; Germany.\\
% ...
$^{105}$CPPM, Aix-Marseille Universit\'e, CNRS/IN2P3, Marseille; France.\\
% ...
$^{120}$$^{(a)}$New York University Abu Dhabi, Abu Dhabi;$^{(b)}$United Arab Emirates University, Al Ain; United Arab Emirates.\\
% ...
$^{124}$Homer L. Dodge Department of Physics and Astronomy, University of Oklahoma, Norman OK; United States of America.\\
% ...
$^{158}$Raymond and Beverly Sackler School of Physics and Astronomy, Tel Aviv University, Tel Aviv; Israel.\\
% ...
$^{a}$ Also Affiliated with an institute covered by a cooperation agreement with CERN.\\
$^{b}$ Also at An-Najah National University, Nablus; Palestine.\\
$^{c}$ Also at Borough of Manhattan Community College, City University of New York, New York NY; United States of America.\\
$^{d}$ Also at Center for High Energy Physics, Peking University; China.\\
$^{e}$ Also at Center for Interdisciplinary Research and Innovation (CIRI-AUTH), Thessaloniki; Greece.\\
$^{f}$ Also at CERN, Geneva; Switzerland.\\
% ...

\end{flushleft}


%

%
%

%

%
%
%
%

\end{document}